\documentclass[usenatbib]{mn2e}
\usepackage{graphicx,times}
\usepackage{bm,url}
\newcommand{\EQ}{\begin{equation}}
\newcommand{\EN}{\end{equation}}
\newcommand{\EQA}{\begin{eqnarray}}
\newcommand{\ENA}{\end{eqnarray}}
\newcommand{\eq}[1]{(\ref{#1})}

\newcommand{\Eq}[1]{Eq.~(\ref{#1})}

\newcommand{\intg}{\int_{-\infty}^{\infty}}
\newcommand{\fjk}{F_{jk}}

\newcommand{\nab}{\nabla}

\newcommand{\flucf}{\bm f}
\newcommand{\meanF}{\overline {\bm F}}

\newcommand{\meanB}{\overline{B}}

\newcommand{\meanemf}{{\cal E}}
\newcommand{\meanEMF}{\mbox{\boldmath ${\cal E}$}}
\newcommand{\tildemeanEMF}{\tilde{\mbox{\boldmath ${\cal E}$}}}
\newcommand{\meanBB}{\bm{\overline{{B}}}}

\newcommand{\meanUU}{\bm{\overline{{U}}}}

\newcommand{\ubkq}{{\overline {\tiluu_{j}(\bm{k}-\bm{q})\tilbb_{k}(\bm{q})}}}
\newcommand{\ubkqt}{{\overline {\tiluu_{j}(\bm{k}-\bm{q},\Omega-\omega)\tilbb_{k}(\bm{q},\omega
)}}}
\newcommand{\uukqt}{{\overline {\tiluu_{j}(\bm{k}-\bm{q},\Omega-\omega)\tiluu_{k}(\bm{q},\omega
)}}}
\newcommand{\uukq}{{\overline {\tiluu_{j}(\bm{k}-\bm{q})\tiluu_{k}(\bm{q})}}}
\newcommand{\fjfk}{{\overline {\tilff_{j}(\bm{k}-\bm{q})\tilff_{k}(\bm{q})}}}
\newcommand{\fjbk}{{\overline {\tilff_{j}(\bm{k}-\bm{q})\tilbb_{k}(\bm{q})}}}
\newcommand{\fjbkt}{{\overline {\tilff_{j}(\bm{k}-\bm{q},\Omega-\omega)\tilbb_{k}(\bm{q},\omega)}}}
\newcommand{\bjbk}{{\overline {\tilbb_{j}(\bm{k}-\bm{q})\tilbb_{k}(\bm{q})}}}
\newcommand{\bjbkt}{{\overline {\tilbb_{j}(\bm{k}-\bm{q},\Omega-\omega)\tilbb_{k}(\bm{q},\omega)}}}
\newcommand{\ujpsi}{{\overline{\tiluu_{j}(\bm{k}-\bm{q})\tilpsi_{i}(\bm{q})}}}
\newcommand{\ulpsil}{{\overline{\tiluu_{l}(\bm{k}-\bm{q})\tilpsi_{l}(\bm{q})}}}
\newcommand{\vpa}{\bm a}
\newcommand{\xx}{\bm{x}}
\newcommand{\qq}{{\bm{q}}}
\newcommand{\pp}{{\bm{p}}}
\newcommand{\kk}{{\bm{k}}}

\newcommand{\UU}{\bm U}
\newcommand{\uu}{\bm{{{u}}}}

\newcommand{\ff}{\bm{{{f}}}}
\newcommand{\BB}{\bm{{{B}}}}

\newcommand{\jj}{{\bm{j}}}

\newcommand{\bb}{\bm{{{b}}}}

\newcommand{\FF}{{\bm{F}}}
\newcommand{\TT}{\bm T}
\newcommand{\GG}{\bm G}
\newcommand{\EE}{\bm E}
\newcommand{\tilbb}{\tilde{b}}
\newcommand{\tiluu}{\tilde{u}}
\newcommand{\tilff}{\tilde{f}}
\newcommand{\tilpsi}{\tilde{\psi}}
\def\ii{{\rm i}}

\def\Rm{R_\mathrm{m}}
\def\Rey{\mbox{\rm Re}}

\def\half{{\textstyle{1\over2}}}

\newcommand{\diff}{\mathrm{d}}
\newcommand{\peff}{p_\mathrm{eff}}

\title[Kinetic and magnetic alpha effects]%
{Kinetic and magnetic alpha effects in nonlinear dynamo theory}
\author[S.~Sur, K.~Subramanian, A.~Brandenburg]%
{Sharanya Sur$^{1}$, Kandaswamy Subramanian$^{1}$ and Axel Brandenburg$^2$ 
\thanks{E-mail: sur@iucaa.ernet.in (SS); kandu@iucaa.ernet.in (KS);
brandenb@nordita.dk (AB)}\\
$^{1}$Inter-University Centre for Astronomy and
        Astrophysics,  Post Bag 4, Ganeshkhind, Pune 411 007, India\\
$^2$NORDITA, AlbaNova University Center, SE - 106 91 Stockholm, Sweden}

\date{}
\begin{document}

\pagerange{\pageref{firstpage}--\pageref{lastpage}} \pubyear{2006}

\maketitle

\begin{abstract}
The backreaction of the Lorentz force on the $\alpha$-effect 
is studied in the limit of small magnetic and fluid Reynolds numbers,
using the first order smoothing approximation (FOSA) to solve both
the induction and momentum equations. Both steady and time dependent forcings
are considered. In the low Reynolds number limit, 
the velocity and magnetic fields can be expressed 
explicitly in terms of the forcing function. 
The nonlinear $\alpha$-effect is then shown to be expressible in several
equivalent forms in agreement with formalisms that are used in
various closure schemes.
On the one hand, one can express $\alpha$ completely
in terms of the helical properties of the velocity field
as in traditional FOSA, 
or, alternatively, as the sum of two terms, 
a so-called kinetic $\alpha$-effect
and an oppositely signed term proportional to the 
helical part of the small scale magnetic field. 
These results hold for both steady and time dependent forcing
at arbitrary strength of the mean field.
In addition, the $\tau$-approximation is considered in the limit
of small fluid and magnetic Reynolds numbers.
In this limit, the $\tau$ closure term is absent and the viscous
and resistive terms must be fully included.
The underlying equations are then identical to those used under FOSA,
but they reveal interesting differences between the steady
and time dependent forcing. 
For steady forcing, the correlation between the
forcing function and the small-scale magnetic field
turns out to contribute in a
crucial manner to determine the net $\alpha$-effect.
However for delta-correlated time-dependent forcing,
this force--field correlation vanishes,
enabling one to write $\alpha$ exactly as the sum of 
kinetic and magnetic $\alpha$-effects, similar to
what one obtains also in the large Reynolds number regime
in the $\tau$-approximation closure hypothesis.
In the limit of strong imposed fields, $B_0$, we find $\alpha \propto
B_0^{-2}$ for delta-correlated forcing, in contrast to the well-known
$\alpha \propto B_0^{-3}$ behaviour for the case of a steady forcing.
The analysis presented here is also shown to be in agreement with
numerical simulations of steady as well as random helical flows.
\end{abstract}
\label{firstpage}

\section{Introduction}

Turbulent mean field dynamos are thought to be at the
heart of magnetic field generation and maintenance in
most astrophysical bodies, like the sun or the galaxy.
A particularly important driver of the mean field dynamo (MFD)
is the $\alpha$-effect which, in the kinematic regime,
depends only on the helical properties of the turbulence.
It is crucial to understand how the $\alpha$-effect gets
modified due to the backreaction of
the generated mean and fluctuating fields.

Using closure schemes or the quasi-linear approximation
it has been argued that, due to Lorentz forces, the $\alpha$-effect gets
``renormalized'' by the addition of a term proportional to the
current helicity of the generated small scale magnetic fields
\citep{pouq,GD,KRR95,S99,KR99,BF,RKR,BS05}.
The presence of such an additional term is uncontroversial if a helical
small scale magnetic field is present even in the absence of a mean field.
However, it has been argued that, in the absence of
such a pre-existing small scale magnetic field, the $\alpha$-effect
can be expressed exclusively in terms of
the velocity field, albeit one which is a solution
of the full momentum equation including the Lorentz force
\citep{proctor,RR07}.
In the latter case, it is not obvious that the helicity of the
small scale magnetic field plays any explicit role in the backreaction to
$\alpha$.
It is important to clarify this issue, as it will decide
how one should understand the saturation of turbulent dynamos,
as well as the possibility of catastrophic quenching
of the $\alpha$-effect and ways to alleviate such quenching.
Here and below, ``catastrophic'' means that $\alpha$ is quenched down
to values on the order of the inverse magnetic Reynolds number.

In order to clarify these conflicting views,
we examine here an exactly solvable model of
the nonlinear backreaction to the $\alpha$-effect,
where we assume small magnetic and fluid Reynolds numbers.
Obviously, this approach does not allow us to address the question
of catastrophic quenching of astrophysical dynamos directly, but it
allows us to make novel and unambiguous statements that help clarifying
the nature of magnetic saturation.
We will show that, at least
in this simple context, both the above viewpoints are
consistent, if interpreted properly. 

\section{Mean field electrodynamics } \label{MFED}

In mean field electrodynamics \citep{KR80,M78}, 
any field $\FF$ is split into a mean field $\meanF$ and a `fluctuating'
small scale field $\flucf$, such that $\FF=\meanF+\flucf$. 
The fluctuating velocity (or magnetic) field is assumed to possess
a correlation length $l$ small compared 
to the length scale $L$ of the variation of the 
mean field.  The magnetic field obeys
the induction equation, 
\EQ
\label{indB}
{\partial{\BB}\over\partial t}=
\eta\nab^2\BB+\nab\times(\UU\times\BB), \quad
\nab\cdot\BB=0,
\EN
where $\UU$ represents the fluid velocity,
$\eta=(\mu_0\sigma)^{-1}$ is the magnetic
diffusivity (assumed constant), $\sigma$ is the electric conductivity,
and $\mu_0$ is the vacuum permeability.
Averaging Eq.~(\ref{indB}), we obtain
the standard mean-field dynamo equation
\EQ
\label{indmeanB}
{\partial{\meanBB}\over\partial t}=
\eta\nab^2\meanBB+\nab\times(\meanUU\times\meanBB+\meanEMF), \quad
\nab\cdot\meanBB=0.
\EN
This averaged equation now has a new term, 
the mean electromotive force (emf) 
$\meanEMF={\overline {\uu\times\bb}}$, which crucially depends on 
the statistical properties of the $\uu$ and $\bb$ fields. The central 
closure problem in mean field theory is to find an expression for 
the correlator $\meanEMF$ in terms of the mean fields.

To find an expression for $\meanEMF$, one needs the
evolution equations for both the fluctuating magnetic field 
$\bb$ and the fluctuating velocity field $\uu$. The first follows from 
subtracting Eq.~(\ref{indmeanB}) from Eq.~(\ref{indB}),
\EQ
\label{flucb}
{\partial{\bb}\over\partial t}=
{\eta\nab^2{\bb}}+\nab\times(\meanUU\times\bb+\uu\times\meanBB)+\GG, \quad
\nab\cdot{\bb}=0.
\EN
Here $\GG=\nab\times(\uu\times\bb)'$ with
$(\uu\times\bb)'=\uu\times\bb-{\overline {\uu\times\bb}}$.
In what follows, we will set the mean field velocity to zero, i.e.\
$\meanUU = 0$ and focus solely on the effect of the
fluctuating velocity.

The evolution equation for $\uu$ can be derived in a similar manner
by subtracting the averaged momentum equation from the full momentum
equation.
We assume the flow to be incompressible with ${\nab\cdot\uu}=0$. 
We get
\EQA
{\partial{\uu}\over\partial t} &=&  
-{1\over\rho}\nab\left(p+{1\over\mu_0}{\meanBB\cdot\bb}\right)+\nu\nab^2\uu 
\nonumber \\
&&+{1\over\mu_0\rho}\left[{(\meanBB\cdot\nab)\bb}
+{(\bb\cdot\nab){\meanBB}}\right]+\flucf+\TT.
\label{flucu}
\ENA
Here $\rho$ is the mass density, $p$ is the perturbed 
fluid pressure, $\nu$ is the kinematic viscosity taken to be constant,
$\flucf$ is the fluctuating force, and
\EQ
{\TT}=-({\uu\cdot\nab\uu})'-{1\over\mu_0\rho}
\left[(\bb\cdot\nab\bb)'-\half\nab(\bb^{2})'\right]
\EN
contains the second order 
terms in ${\uu}$ and ${\bb}$. 
Here, primed quantities indicate deviations from the mean,
i.e.\ $X' = X - \overline{X}$.
We will also redefine $\bb/\sqrt{\mu_0\rho} \to \bb$, 
by setting $\mu_0\rho=1$, so that
the magnetic field is measured in velocity units.

In order to find $\meanEMF$ under the influence of the Lorentz force
one has to solve Eqs~(\ref{flucb}) and (\ref{flucu}) simultaneously
and compute ${\overline {\uu\times\bb}}$. In general this is a
difficult problem and one has to take recourse to closure 
approximations or numerical simulations. To make progress 
we assume here $\Rm = ul/\eta \ll 1$ and $\Rey= ul/\nu \ll 1$; that is 
both the magnetic and fluid Reynolds numbers are small
compared to unity.
In this case there is no small scale dynamo action and so the small scale
magnetic field is solely due to shredding the large scale magnetic field.
Here $u$ and $b$ (see below) are typical strengths of the
fluctuating velocity and magnetic fields respectively.
In the low magnetic Reynolds number limit the ratio of the 
first nonlinear term in $\GG$ to the resistive term in Eq.~(\ref{flucb})
is $\sim (ub/l)/(\eta b/l^2) \sim \Rm \ll 1$. So this part of $\GG$ can be neglected
compared to the resistive term. 
(Note that the second term in $\GG$ vanishes automatically when
taking the averages to evaluate the mean emf.)
Neglecting the nonlinear term, the generation rate of
$b$ is $\sim u \meanB/l$, while
its destruction rate is $\sim \eta b/l^2$. 
Equating these two rates, this also implies that $b \sim \Rm \meanB$
and the fluctuation field is only a small perturbation to mean fields.
Similarly the ratio of the nonlinear advection term 
to the viscous term in Eq.~(\ref{flucu}), 
is $\sim (u^2/l)/(\nu u/l^2) \sim \Rey\ll 1$ and the ratio of the parts of the
Lorentz force nonlinear in $\bb$ to that linear in $\bb$ is 
$\sim (b^2/l)/(b{\overline B}/l) \sim \Rm \ll 1$. So $\TT$ can also be
neglected in Eq.~(\ref{flucu}).
In this limit, one can therefore apply the
well known first order smoothing approximation (FOSA).
It is sometimes also referred to as the second order correlation approximation
\citep[or SOCA; see, e.g.,][]{KR80}.
This approximation consists of neglecting the nonlinear terms $\GG$ and $\TT$,
to solve {\it both} the induction and momentum equation.
Since FOSA is applied to the momentum equation as well, we
will refer to this as ``double FOSA''.
In order to make the problem analytically tractable,
we will take ${\meanBB=\BB_{0}}=\mbox{const}$.
This also allows us to isolate the $\alpha$-effect in a straightforward fashion.
In the next section we begin by considering for simplicity the case of steady
forcing.
It is then possible to also neglect the time derivatives in
Eqs~(\ref{flucb}) and (\ref{flucu}).
We return to consider time dependent
forcing in detail in Section~\ref{Section:timedep}.

\section{Computing $\meanEMF$ for steady forcing} 
\label{Section:meanemf}
Under the assumptions highlighted above, one can solve directly for 
${\uu}$ and ${\bb}$ in terms of the forcing function ${\flucf}$. 
This in turn allows the calculation of the mean emf in four ways.
\begin{itemize}
\item[A.] We use the induction equation to solve for $\bb$ in terms of $\uu$.
Then one can write the emf completely in terms of the velocity field, 
as in normal FOSA and then substitute for ${\uu}$ in terms of ${\flucf}$.
\item[B.] We compute $\meanEMF={\overline {\uu\times\bb}}$ directly.
\item[C.] We use the momentum equation to solve for ${\uu}$ in terms 
of ${\bb}$ and the forcing function ${\flucf}$, and then substitute 
for ${\bb}$ in terms of ${\flucf}$.
\item[D.] Compute $\meanEMF$ from the
$\partial\meanEMF/\partial t=0$ relation, as in $\tau$-approximation closures.
\end{itemize}
We will show that all four methods give the same answer for the mean emf
in terms of the forcing function $\flucf$. The first Method~A gives
the traditional FOSA result for the $\alpha$-effect being
dependent on the helical properties of the velocity field,
while Method~C can be interpreted to reflect the idea of 
a renormalized $\alpha$ due to the helicity of small scale magnetic fields.
But we show that the final answer in terms of the forcing is identical.

Before going into the various methods as highlighted above, 
we solve for ${\uu}$ and ${\bb}$ in terms of the forcing function 
${\flucf}$.
In the low conductivity limit, neglecting the time variation of $\bb$ 
in Eq.~(\ref{flucb}) we have, 
\EQ
-{\eta\nab^2{\bb}}=\BB_0\cdot\nab\uu.
\label{loconflucb}
\EN
Similarly, in the limit of low $\Rey$ and $\Rm$, Eq.~(\ref{flucu}) becomes, 
\EQ
-{\nu\nab^2\uu}=
\BB_0\cdot\nab{\bb}+\flucf-\nab\peff,
\label{loconflucu}
\EN
where $\peff$ combines the hydrodynamic and the magnetic pressure.
Using the incompressibility condition, one can eliminate
$\peff$.
We will solve these equations in Fourier space.
Throughout this paper we will be using the convention
\EQ
{\tilde{\uu}}({\bm k})=
{1\over (2\pi)^3}{\int{\bm u}({\bm x})e^{-\ii\kk\cdot\xx}\diff{\bm x}},
\EN
which satisfies the inverse relation 
\EQ
{\uu}({\bm x})=
{\int{{\tilde{\uu}}({\bm k})e^{\ii\kk\cdot\xx}\diff{\bm k}}}.
\EN
In Fourier space, Eqs~(\ref{loconflucb})  and (\ref{loconflucu}) become
\EQ
\label{bintu}
{\eta k^2\tilbb_{i}}({\bm k})=
(\ii\kk\cdot\BB_0)\,\tiluu_{i}(\kk),
\EN
\EQ
\label{uintbf1}
\nu k^2\tiluu_{i}(\kk)=
(\ii\kk\cdot\BB_0)\,\tilbb_{i}(\kk)+\tilff_{i}(\kk),
\EN
where we have chosen the forcing to be divergence free,
with $\ii\kk\cdot\tilde{\bm{f}}=0$.
We can therefore solve the above two equations simultaneously 
to express $\tilde{\uu}$ and $\tilde{\bb}$ completely in terms of $\tilde{\flucf}$,
\EQ
\label{uintf}
{\tiluu_{i}}({\bm k})=
\frac{\tilff_i(\kk)}{\nu k^2+({\BB_0\cdot\kk})^2/\eta k^2},
\EN
\EQ
\label{bintf}
{\tilbb_i}(\kk)=
\frac{\tilff_i(\kk)}{\nu k^2+({\BB_0\cdot\kk})^2/\eta k^2}
\,\frac{\ii\kk\cdot\BB_0}{\eta k^2}.
\EN
We can use these solutions to calculate $\meanEMF$. For getting
an explicit expression, we also need the equal time
force correlation function.
For isotropic and homogeneous forcing, this is given by
\EQ
\label{ffcor}
\overline{{\tilff_{j}}({\pp},t)\;{\tilff_{k}}({\qq},t)} =
{\delta^3({\pp+\qq})}{{\fjk}({\qq})}.
\EN
Here, $\fjk$ is the force spectrum tensor which is given by
\EQ
\label{fjk}
{\fjk}(\kk)=
P_{jk}\frac{\Phi(k)}{4\pi k^{2}} +
\epsilon_{jkm}\frac{{\ii k_{m}}{\chi(k)}}{8{\pi}k^{4}},
\EN
where ${P_{jk}}={\delta_{jk}}-{{k_{j}}{k_{k}}/k^2}$ is the 
projection operator,
and $\Phi(k)$ and $\chi(k)$ are spectra characterizing the
mean squared value and the helicity of the forcing function,
normalized such that
\EQ
\int_0^\infty\Phi(k)\;\diff{k}=\half\overline{\ff^2}
\equiv\half A_{\rm f}^2.
\EN
\EQ
\int_0^\infty\chi(k)\;\diff{k}=\overline{\ff\cdot(\nab\times\ff)}
\equiv H_{\rm f}.
\EN
The mean emf can be written as
\EQ
\label{meanemf}
\meanemf_{i}({\bm x})=
{\epsilon_{ijk}}{\overline {{\uu}_j({\bm x}){\bb}_k({\bm x})}}
= \int{{\tilde{\meanemf_{i}}({\bm k})}\;e^{\ii\kk\cdot\xx}
\diff{\bm k}},
\EN
where the Fourier transform $\tildemeanEMF$ is given by
\EQ
\label{fsemf}
{\tilde{\meanemf_{i}}({\bm k})}= \epsilon_{ijk} {\int{\ubkq\;\diff{\bm q}}}.
\EN
We now turn to the calculation of the nonlinear mean emf and the resulting
nonlinear $\alpha$-effect in the four different methods outlined above.

\subsection{Method~A: express $\bb$ in terms of $\uu$ and then 
solve for $\meanEMF$}
\label{metha}
In this approach we use the induction equation to solve for
$\bb$ in terms of $\uu$.
Using \Eq{bintu} to express $\bb$ in terms of $\uu$ in \Eq{fsemf} gives
\EQ
\label{emfbu1}
{\tilde{\meanemf_{i}}({\bm k})}=
{\ii}{\epsilon_{ijk}}\;{\int{\frac{{\BB_0\cdot\qq}}{\eta q^2}
\;\uukq\;\diff{\bm q}}}.
\EN
At this stage one can put the emf completely in terms of 
the velocity field and recover the 
usual FOSA expression that in the low conductivity and isotropic limit, 
the ${\alpha}$-effect is related to
the helicity of the velocity potential
\citep{KR80,RR07}.
This can be shown in the following manner: since $\nab\cdot\uu =0$,
the velocity field can be expressed as 
$\uu=\nab\times\bm{\psi}$, where $\bm{\psi}$ is the velocity vector
potential with the gauge condition
$\nab\cdot\bm{\psi}=0$. We then have
$\tiluu_k(\qq)=
{\ii}q_p\epsilon_{kpl}\tilpsi_l(\qq)$.
Substituting this expression in Eq.~(\ref{emfbu1}),
and using the fact that the velocity field is divergence-less, 
we get 
\EQA
\label{emfbu2}
\nonumber
{\tilde{\meanemf_{i}}({\bm k})}&=&
k_j{\int{\frac{{\BB_0\cdot\qq}}{\eta q^2}\;
\ujpsi\;\diff{\bm q}}} \\
&&-{\int{\frac{{\BB_0\cdot\qq}}{\eta q^2}\;
q_i\;\ulpsil\;\diff{\bm q}}}.
\ENA
For homogeneous and isotropic turbulence, the $\ujpsi$
correlation is proportional to $\delta^3(\kk)$. Since the first term in
Eq.~(\ref{emfbu2}) is $\propto k_j$, it 
does not contribute to $\meanEMF$. Therefore,
\EQ
\label{emfbu3}
{\tilde{\meanemf_{i}}({\bm k})}=
-{\int{\frac{{B_{0m}\;q_m}}{\eta q^2}\;
q_i\;\ulpsil\;\diff{\bm q}}}.
\EN
Again, for homogeneous and isotropic turbulence,
the $\ulpsil$ correlation is $\propto \delta^3(\kk) g(|\qq|)$.
One can then carry out the angular integral in 
Eq.~(\ref{emfbu3}) using 
$\int (q_m q_i/q^2) (d\Omega/4\pi)
=(1/3) \delta_{mi}$ to get
$ {\meanemf_{i}(\xx)}= \alpha B_{0i}$,
where $\alpha$ is given by 
\EQ
\label{uualpha}
\alpha=-\frac{1}{3\eta}\,\overline{\bm{\psi}\cdot\uu},
\EN
which is identical to the expressions obtained by \cite{KR80};
see also \cite{RR07}.
(Note that when the Lorentz force becomes important the
assumption of isotropy in the above derivation breaks down and 
the $\alpha$-effect becomes anisotropic,
as calculated below and detailed in Section 3.5.)

Since we have already solved for the velocity field explicitly, 
we can now derive an expression for the mean emf in this 
approach. Substituting the velocity
in terms of the forcing function, the mean emf in coordinate 
space is given by
\EQ
\label{emfmeta}
{\meanemf_{i}({\bm x})}=
\ii{\epsilon_{ijk}}\int{{\BB_{0}\cdot{\bm q}}\over{\eta q^2}}
\frac{F_{jk}(\qq)}
{\left[\nu q^2 +{(\BB_ 0\cdot\qq)^2}/{\eta q^2}\right]^2}\;\diff{\qq}.
\EN
Here $F_{jk}$ the spectrum tensor for the force-field given 
by Eq.~(\ref{fjk}). Note that only the antisymmetric part of
$F_{jk}$ contributes to $\meanEMF$, due to the presence of
$\epsilon_{ijk}$ on the RHS of the above equation.
We can also write $\meanEMF$ as
\EQ
\label{emfmetan}
{\meanemf_{i}({\bm x})}=
\ii{\epsilon_{ijk}}\int\frac{\BB_0\cdot\qq}{(\eta q^2)(\nu q^2)^2}
\frac{F_{jk}(\qq)}{\left[1+N\right]^2}\;\diff{\qq},
\EN
where $N=({\BB_0\cdot\qq})^2/(\eta\nu q^4)$ determines the importance of the 
Lorentz forces on the mean emf. It is to be noted that the limit of
small Lorentz forces corresponds to taking $N \ll 1$ above.
\subsection{Method~B: compute $\meanEMF$ directly}
\label{methb}
In this approach, we directly compute 
$\meanEMF={\overline {\uu\times\bb}}$ by
substituting $\uu$ and $\bb$ 
in terms of $\flucf$, using Eqs~(\ref{uintf}) and (\ref{bintf}).
We then get
\EQ
\label{meanemfff}
\tilde{\meanemf_i}(\kk)=
\ii\epsilon_{ijk}\; \int \frac{\BB_{0}\cdot\qq}{\eta q^2}
\; \frac{\fjfk}{\gamma(\kk-\qq)\gamma(\qq)} \;\diff{\bm q},
\EN
where we have defined
$\gamma(\qq) = \nu q^2+({\BB_0\cdot\qq})^2/\eta q^2$.
Substituting for the force correlation, the mean emf in
coordinate space is given by 
\EQ
\label{emfmetb}
{\meanemf_{i}({\bm x})}=
\ii{\epsilon_{ijk}}\int{{\BB_{0}\cdot{\bm q}}\over{\eta q^2}}
\frac{F_{jk}(\qq)}
{\left[\nu q^2 +{(\BB_ 0\cdot\qq)^2}/{\eta q^2}\right]^2}\;\diff{\qq},
\EN
which is identical to Eq.~(\ref{emfmeta}) for $\meanEMF$
obtained by Method~A. 

\subsection{Method~C: express $\uu$ in terms of $\bb$ and 
then solve for $\meanEMF$} 
\label{methc}
Note that one could also start from the momentum equation
to compute $\meanEMF$. In this approach, we first solve for $\uu$ in terms 
of $\bb$ and the forcing function $\flucf$, 
and then substitute for $\bb$ in terms of $\flucf$ 
using Eq.~(\ref{bintf}).
The difference from the earlier treatments 
will be an additional term containing an
$\flucf\times\bb$-like correlation, which turns out to be
essential for calculating the $\meanEMF$ correctly. 
Using Eq.~(\ref{uintbf1}) one can write 
\EQ
\label{uintbf2}
{\tiluu_{i}}({\bm k})=
\frac{1}{\nu k^2}\left[(\ii\kk\cdot\BB_{0})\,{\tilbb_{i}}({\bm k})
+{\tilff_{i}}({\bm k})\right].
\EN
From Eq.~(\ref{fsemf}) the mean emf can then be written as  
\EQA
\nonumber
{\tilde{\meanemf_{i}}({\bm k})}&\!=\!&
{\epsilon_{ijk}}{\int{\frac{1}{\nu(\kk-\qq)^2}\fjbk\;\diff{\bm q}}} \\
&&\!\!+{\ii}{\epsilon_{ijk}}{\int\frac{\BB_0\cdot({\bm{k-q}})}{\nu(\kk-\qq)^2}
\,\bjbk\;\diff{\bm q}}.
\label{metC}
\ENA
Here the first term involves the $\flucf\times\bb$-like correlation.
To elucidate the meaning of the second term it is useful
to define the magnetic field as
$\bb = \nab\times\vpa$, where $\vpa$ is the small scale
magnetic vector potential in the Coulomb gauge ($\nab\cdot\vpa =0$).
Then, for isotropic small scale fields, following the approach 
in Method~A, the second term in Eq.~(\ref{metC}) 
gives a contribution to $\meanEMF$ of the form $\hat{\alpha}_{\rm M}\BB_{0}$, where
\EQ
\hat{\alpha}_{\rm M}={1\over3\nu}\,\overline{\vpa\cdot\bb}.
\EN
So this contribution is proportional to the magnetic helicity
of the small scale magnetic field (analogous to the helicity of the vector
potential of the velocity field).

If we substitute $\bb$ in terms of $\flucf$ from Eq.~(\ref{bintf})
and then integrate over the delta function, the mean emf in coordinate 
space can be expressed as
\EQA
\label{emfmc}
{\meanemf_{i}({\bm x})}&=&
\ii{\epsilon_{ijk}}\int\frac{\BB_0\cdot\qq}{\eta q^2}
\frac{F_{jk}(\qq)}{(\nu q^2)^2\;\left[1+N\right]}\diff{\qq} \\ \nonumber
&&-\ii{\epsilon_{ijk}}\int\frac{\BB_0\cdot\qq}{\eta q^2}
\frac{F_{jk}(\qq)}{(\nu q^2)^2\;\left[1+N\right]}
\,\frac{N}{1+N}\,\diff{\qq}.
\ENA
The two terms on the RHS of the above equation have an 
interesting interpretation. As mentioned above, the limit of small
Lorentz forces corresponds to taking 
$N = ({\BB_0\cdot\qq})^2/\eta\nu q^4 \ll 1$.
In this limit the second integral vanishes while the first one
[i.e.\ the $\flucf\times\bb$-like correlation in Eq.~(\ref{emfmc}),
which is really a $(\nabla^{-2}\flucf)\times\bb$ correlation]
goes over to a kinematic $\alpha$-effect.
[One can see by comparing Eq.~(\ref{emfmetan}) and the first term in
Eq.~(\ref{emfmc}) that the two are identical in the
limit $N \ll 1$].
In fact, this part of the $\alpha$-effect can be obtained from
Eqs~(\ref{uintf}) and (\ref{bintf}) by neglecting the
Lorentz force in the expression for ${\tiluu_{i}}({\bm k})$.
In the following we refer to the contribution of the
field-aligned component of this term 
divided by $B_0$, as the $\hat{\alpha}_{\rm F}$ term,
because it comes from the $\flucf\times\bb$-like correlation.

As $N$ is increased the contribution from the first term decreases.
In the same limit, the second term,
which depends on the magnetic helicity, gains in
importance. Since it has the opposite sign, it partially cancels the 
first term and further suppresses the total $\alpha$-effect.
This is reminiscent of the suppression of
the kinetic alpha due to the addition of a magnetic alpha
(proportional to helical part of $\bb$) found in several
closure models \citep{pouq,KR82,GD,BF02,BS05b}.

When adding the two terms in Eq.~(\ref{emfmc}),
the mean emf turns out to be 
\EQ
\label{emfmetc}
{\meanemf_{i}({\bm x})}=
\ii{\epsilon_{ijk}}\int{{\BB_{0}\cdot{\bm q}}\over{\eta q^2}}
\frac{F_{jk}(q)}
{\left[\nu q^2 +{(\BB_ 0\cdot\qq)^2}/{\eta q^2}\right]^2}\;\diff{\qq},
\EN
which is identical to the expressions obtained in Methods~A and B;
see Eqs~(\ref{emfmeta}) and (\ref{emfmetb}), respectively.

\subsection{Method~D: compute $\meanEMF$ from the
$\partial\meanEMF/\partial t=0$ relation}

In recent years the so-called $\tau$-approximation has received increased
attention \citep{KMR96,BF02,RKR,BS05,RR07}.
This involves invoking a closure whereby triple correlations 
which arise during the evaluation of $\partial\meanEMF/\partial t$,
are assumed to provide a damping term proportional to $\meanEMF$ itself.
In the present context there is no need to invoke a closure for the
triple correlations, because
these terms are small for low fluid and magnetic Reynolds numbers.
It turns out that the correct expression for $\meanEMF$ can still be
derived in the same framework, where one evaluates the
$\partial\meanEMF/\partial t$ expression.

The expression for $\partial\meanEMF/\partial t$ 
is governed by two terms, $\overline{\dot{\uu}\times\bb}$ and
$\overline{\uu\times\dot{\bb}}$, where dots denote partial time
differentiation.
Of course, both $\dot{\uu}$ an $\dot{\bb}$ vanish in the present case,
but this is the result of a cancellation of driving and dissipating terms.
In the present analysis both terms will be retained, because the
dissipating term, which is related to the desired $\meanEMF$, can then
just be written as the negative of the driving term.

We perform the analysis in Fourier space and begin by defining
$\EE(\kk,\qq) = \overline{\tilde{\uu}(\kk-\qq)\times\tilde{\bb}(\qq)}$.
Note the required 
\EQ
\meanEMF = \int {\EE(\kk,\qq)} e^{ i {\kk\cdot\xx}}
\diff{\bm k}\, \diff{\bm q}.
\label{EEdef}
\EN
To calculate the time derivative $\partial\meanEMF/\partial t$, one
needs to evaluate $\dot{\EE} = \dot{\EE}_{\rm K} + \dot{\EE}_{\rm M}$, where
\EQ
\dot{\EE}_{\rm K}(\kk,\qq) = 
\overline{\tilde{\uu}(\kk-\qq)\times\dot{\tilde{\bb}}(\qq)},
\label{ubdottindep}
\EN
\EQ
\dot{\EE}_{\rm M}(\kk,\qq) = 
\overline{\dot{\tilde{\uu}}(\kk-\qq)\times\tilde{\bb}(\qq)}.
\label{udotbtindep}
\EN
For $\dot{\tilde{\uu}}$ and $\dot{\tilde{\bb}}$ we restore the
time derivatives in Eqs~(\ref{bintu}) and (\ref{uintbf1}), 
and obtain 
\EQA
\label{emfKdot}
\dot{\EE}_{\rm K}&=&\ii{\qq\cdot\BB_0}\,
[\overline{\tilde{\uu}(\kk-\qq)\times\tilde{\uu}(\qq)}]
-\eta q^2 {\EE},
\\
\label{emfMdot}
\dot{\EE}_{\rm M} &=&\ii(\kk-\qq)\cdot\BB_0\,
[\overline{\tilde{\bb}(\kk-\qq)\times\tilde{\bb}(\qq)}]
-\nu (\kk-\qq)^2 \EE
\nonumber \\
&&+\overline{\tilde{\ff}(\kk-\qq)\times\tilde{\bb}(\qq)}.
\ENA
Since all time derivatives are negligible,
we can simplify the RHS of the above equations 
by using Eqs~(\ref{uintf}) and (\ref{bintf}) to express $\tilde{\uu}$ and
$\tilde{\bb}$ in terms of the forcing function.
Adding the two Eqs~(\ref{emfKdot}) and (\ref{emfMdot}) yields
\EQA
\left[
\eta q^2 + \nu ({\kk -\qq})^2\right] {{\EE}_{i}} = 
\delta^3(\kk) \frac{\ii\qq\cdot\BB_0}
{\eta q^2}
\frac{\epsilon_{ijk}\;F_{jk}(\qq)}{\gamma(\qq)\gamma(\kk-\qq)}
\nonumber \\
 \times \left[\eta q^2 + \gamma(\kk -\qq) - \frac{((\kk-\qq)\cdot\BB_0)^2}
{\eta(\kk-\qq)^2} \right],
\label{Ekq}
\ENA
where the function $\gamma$ was defined just below \Eq{meanemfff}.
The expression in the squared brackets on the right hand side
of the expression above exactly reduces to
$\eta q^2 + \nu (\kk -\qq)^2$ and so, in the steady state limit,
$\partial/\partial t=0$, we have
\EQ
\label{tau1}
{{\EE}_{i}(\kk,\qq)} = \delta^3(\kk) \frac{\ii{\qq\cdot\BB_0}}{\eta q^2}
\frac{\epsilon_{ijk}\; F_{jk}(\qq)}{\gamma(\qq)\gamma(\kk-\qq)}.
\EN
Using this expression in Eq.~(\ref{EEdef}), and integrating
over $\kk$, we again recover the form of $\meanEMF$ identical
to Methods~A, B and C.

Thus, in this simple example where one can apply
FOSA to both the induction and
momentum equations, one gets identical expression for
$\meanEMF$ in
terms of the correlation properties of the forcing
function $\flucf$, in all the four methods.

\subsection{The nonlinear $\alpha$-effect}
We now compute the nonlinear $\alpha$-effect explicitly
from the expression of $\meanEMF$ as obtained in 
the four methods discussed above. As has been mentioned
earlier, only the antisymmetric part of $F_{jk}$
contributes to $\meanEMF$ in Eq.~(\ref{fjk}), so Eq.~(\ref{emfmetc}) takes
the simple form
\EQ
\label{alpha1}
{\meanemf_{i}({\bm x})}=
\ii{\epsilon_{ijk}}\!\int{\!{\BB_{0}\cdot{\bm q}}\over{\eta q^2}}
\frac{{\ii}q_m\epsilon_{kjm}\chi(q)}
{8\pi q^4\left[\nu q^2 +{(\BB_ 0\cdot\qq)^2}/{\eta q^2}\right]^2}
\;\diff{\qq},
\EN
Contraction between the two $\epsilon$'s and solving for
$\alpha=\meanEMF\cdot\BB_0/B_0^2$ leads to
\EQ
\label{alpha2}
\alpha=
-\int{\chi(q)\over\eta\nu^2q^6}\,
\frac{({\hat{\BB}_0\cdot\hat{\qq}})^2}
{\left[1+(\hat{\BB}_ 0\cdot\hat{\qq})^2\beta^2\right]^2}\,
{\diff{\qq}\over4\pi q^2},
\EN
where we have introduced $\beta^2=B_0^2/(\eta\nu q^2)$ and
hats denote unit vectors.
The solution involves an angular integral with respect to the cosine
of the polar angle, $\mu=\hat{\BB}_0\cdot\hat{\qq}$,
\EQ
F(\beta)=\int_{-1}^1{\mu^2\,\diff{\mu}\over(1+\beta^2\mu^2)^2}
={1\over\beta^2}\left({\tan^{-1}\!\beta\over\beta}-{1\over1+\beta^2}\right),
\EN
so that
\EQ
\label{alpha3}
\alpha=-{1\over2\eta\nu^2}
\int_{0}^{\infty}\frac{\chi(q)}{q^6}\,F(\beta)\,\diff{q}.
\EN 
Note that for small values of $\beta$ we have $F(\beta)\approx2/3-4\beta^2\!/5$.
In the limit of large values of $B_0$ and $\beta$ we have
$F(\beta)\to\pi/(2\beta^3)$, so the expression of $\alpha$ reduces to
\EQ
\label{alpha5}
\alpha\to
-\frac{\pi}{4B_0^3}\sqrt{\frac{\eta}{\nu}}
\int_{0}^{\infty}\frac{\chi(q)}{q^3}\,\diff{q}
\quad\mbox{(for $B_0\to\infty$)}.
\EN
So, in the asymptotic limit of large $B_0$, we have
$\alpha \rightarrow B_0^{-3}$.
This is a well-known result that goes back to the pioneering
works of \cite{Mof72} and \cite{Rue74}; see also \cite{RK93}.
Note that $\meanEMF\times\BB_0=\bm{0}$, because the corresponding
angular integral would be over the product of a sine and cosine term
which vanishes.

To illustrate further the dependence of
$\alpha$ on $B_0$ we need to adopt some form for
the spectrum $\chi(q)$. We assume that
the forcing is at a particular wavenumber, $q_0$, and choose 
$\chi(q) = H_{\rm f}\delta(q - q_{0})$ where $H_{\rm f}$ is the helicity 
of the forcing.
Then the integration of the delta function simply gives
\EQ
\label{totalpha}
\frac{\alpha}{\alpha_0}=
{3\over2}\left({B_0\over B_{\rm cr}}\right)^{-2}
\left[{\tan^{-1}(B_0/B_{\rm cr})\over B_0/B_{\rm cr}}
-{1\over1+B_0^2/B_{\rm cr}^2}\right]\!,
\EN
where we have defined
\EQ
\alpha_0=-H_{\rm f}/(3\eta\nu^2\!q_0^6),\quad
B_{\rm cr}=\sqrt{\eta\nu}q_{0}.
\EN
If we express $\alpha = \hat{\alpha}_{\rm F} + \hat{\alpha}_{\rm M}$, where 
$\hat{\alpha}_{\rm F}$ is computed from the $(\nabla^{-2}\flucf)\times\bb$ term
and $\hat{\alpha}_{\rm M}$ from the $(\nabla^{-2}\bb)\times\bb$ term
in Eq.~(\ref{emfmc}), we have
\EQ
\label{alphaF}
\frac{\hat{\alpha}_{\rm F}}{\alpha_0}=
3\left({B_0\over B_{\rm cr}}\right)^{-2}
\left[1 - {\tan^{-1}(B_0/B_{\rm cr})\over B_0/B_{\rm cr}}\right],
\EN
\EQ
\frac{\hat{\alpha}_{\rm M}}{\alpha_0}= 
3\left({B_0\over B_{\rm cr}}\right)^{-2}\!
\left[
{3\over 2}{\tan^{-1}(B_0/B_{\rm cr})\over B_0/B_{\rm cr}}
-\frac{3/2+B_0^2/B_{\rm cr}^2}{1+B_0^2/B_{\rm cr}^2}
\right]\!.
\label{alphaM}
\EN
The hats on the $\alpha$s indicate that a special choice has been
made to divide $\alpha$ up into different contributions.
A different choice without hats that had been derived under the $\tau$
approximation, will be discussed in Section~\ref{tauUnsteady}.

\begin{figure} \begin{center}
\includegraphics[width=\columnwidth]{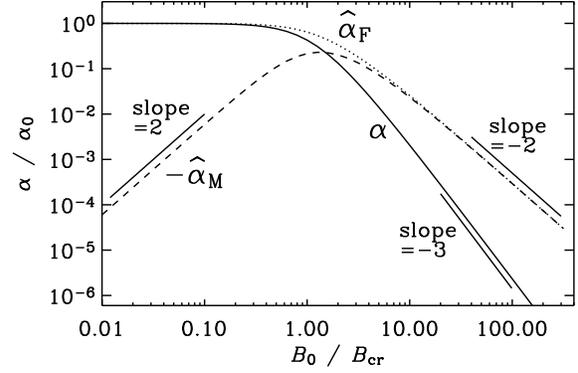}
\end{center}\caption[]{Variation of $\alpha$, $\hat{\alpha}_{\rm F}$, and
$-\hat{\alpha}_{\rm M}$ with $B_0$ from the analytical theory.
Note the mutual approach of $\hat{\alpha}_{\rm F}$ and $-\hat{\alpha}_{\rm M}$
(asymptotic slope $-2$) to produce a lower, quenched value
(asymptotic slope $-3$).}
\label{panalyt_vs_B0}
\end{figure}

We plot in Fig.~\ref{panalyt_vs_B0} the variation of $\alpha$ with $B_0$.
This shows that $\alpha \sim \alpha_0$ for $B_0 \la B_{\rm cr}$
and in the asymptotic limit $\alpha$ decreases $\propto B_0^{-3}$.
This figure also shows the variations of $\hat{\alpha}_{\rm F}$ 
and $\hat{\alpha}_{\rm M}$ with $B_0$ as predicted from Eqs~(\ref{alphaF})
and (\ref{alphaM}).
Both decrease asymptotically like $B_0^{-2}$ because here, unlike in
Eq.~(\ref{totalpha}), the term in squared brackets remains constant.
Their sum decreases as $B_0^{-3}$. 
One could also define a kinetic $\hat{\alpha}_{\rm K}$
from the $(\nabla^{-2}\uu)\times\uu$ term
in Eq.~(\ref{emfmetan}). In this case, for steady forcing we have 
$\alpha = \hat{\alpha}_{\rm K} = \hat{\alpha}_{\rm F}+\hat{\alpha}_{\rm M}$.

\begin{figure} \begin{center}
\includegraphics[width=\columnwidth]{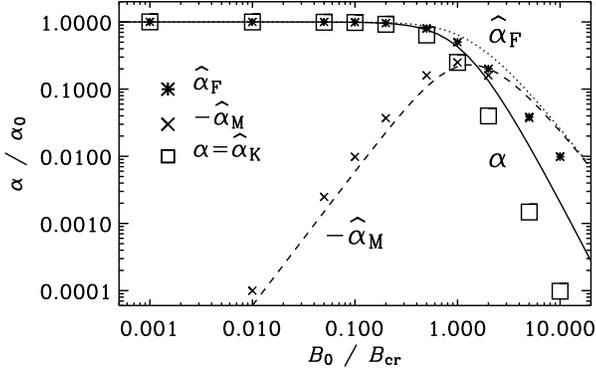}
\end{center}\caption[]{Variation of $\alpha$, $\hat{\alpha}_{\rm F}$, and
$-\hat{\alpha}_{\rm M}$ with $B_0$ for the ABC flow
at low fluid and magnetic Reynolds numbers ($\Rey=R_{\rm m}=10^{-4}$),
compared with the analytic theory predicted for a fully isotropic flow.
Note that the numerically determined values of $\hat{\alpha}_{\rm F}$ are
smaller and those of $-\hat{\alpha}_{\rm M}$ larger than the corresponding
analytic values.
However they still add up to satisfy the relation 
$\hat{\alpha}_{\rm F} + \hat{\alpha}_{\rm M} = \alpha$, predicted by analytic theory.
In all cases the numerically determined values of $\hat{\alpha}_{\rm K}$
agree with the numerically determined values of $\alpha$.
}\label{palpall_vs_B0}
\end{figure}

\subsection{Comparison with simulations}
\label{NumericalVerification}

Simulations allow us to alleviate some of the restrictions imposed by the
analytical approach such as the limit of low fluid and magnetic Reynolds
numbers, but they also introduce additional restrictions related for
example to the degree of anisotropy.
We adopt here a simple and frequently used steady and monochromatic
forcing function that is related to an ABC flow, i.e.\
\EQ
\ff(\xx)={A_{\rm f}\over\sqrt{3}}\pmatrix{
\sin k_{\rm f}z+\cos k_{\rm f}y\cr
\sin k_{\rm f}x+\cos k_{\rm f}z\cr
\sin k_{\rm f}y+\cos k_{\rm f}x},
\EN
where $A_{\rm f}$ denotes the amplitude and $k_{\rm f}$ the
wavenumber of the forcing function.
This forcing function is isotropic with respect to the three
coordinate directions, but not with respect to other directions.
The helicity of this forcing function is $H_{\rm f} = k_{\rm f} A_{\rm f}^2$.
We use the \textsc{Pencil Code},\footnote{
\url{http://www.nordita.dk/software/pencil-code}}
which is a high-order finite-difference code (sixth order in space and
third order in time) for solving the compressible hydromagnetic equations.
We adopt a box size of $(2\pi)^3$
and take $A_{\rm f}=10^{-4}$, $k_{\rm f}=1$, and determine
$\alpha=\meanemf_z/B_{0z}$ as well as $\hat{\alpha}_{\rm K}$,
$\hat{\alpha}_{\rm F}$, and $\hat{\alpha}_{\rm M}$, which are given
respectively by the three integrals in Eqs~(\ref{emfbu1}) and (\ref{metC}).
The result is shown in Fig.~\ref{palpall_vs_B0}.
For these runs a resolution of just $32^3$ meshpoints is sufficient,
as demonstrated by comparing with runs with $64^3$ meshpoints.

For $B_0/B_{\rm cr}\leq1$ the resulting values of $\alpha$
agree in all cases perfectly with both $\hat{\alpha}_{\rm K}$ and
$\hat{\alpha}_{\rm F}+\hat{\alpha}_{\rm M}$.
However, for $B_0/B_{\rm cr}>1$ the numerically determined values
of $\alpha$ are smaller than those expected theoretically using $q_0=k_{\rm f}$.
As in the analytical theory, the quenching is explained by an
uprise of $-\hat{\alpha}_{\rm M}$.
Note, however, that in the simulations this quantity attains
a maximum at somewhat weaker field strength than in the analytical theory;
cf.\ the dashed line and crosses in Fig.~\ref{palpall_vs_B0}.
We believe that this discrepancy is explained by an insufficient
degree of isotropy of the forcing function.

\begin{figure}\begin{center}
\includegraphics[width=\columnwidth]{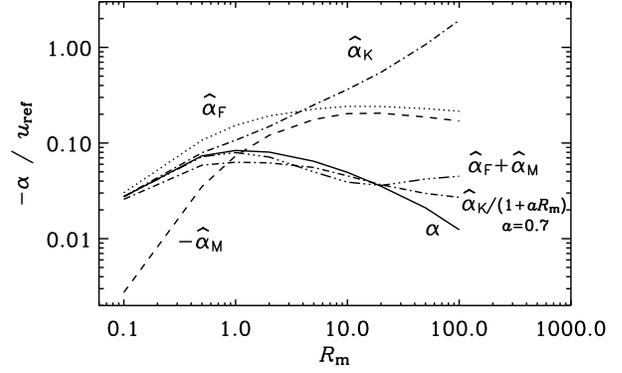}
\end{center}\caption[]{
Dependence of the $\alpha$-effect on $R_{\rm m}$
for fixed field strength, $B_0=u_{\rm ref}$,
for ABC-flow forcing at $\Rey=1$.
Note the agreement between $\alpha$ and $\hat{\alpha}_{\rm F}+\hat{\alpha}_{\rm M}$
as well as $\hat{\alpha}_{\rm K}/(1+aR_{\rm m})$ for $R_{\rm m}\leq30$.
}\label{palptau_vs_Rm_nu2_B2_corr2}\end{figure}

Finally, it is interesting to
address the question of the Reynolds number dependence
of the quenching behaviour.
In Fig.~\ref{palptau_vs_Rm_nu2_B2_corr2} we show the results for
$\alpha$, $\hat{\alpha}_{\rm K}$, $\hat{\alpha}_{\rm F}$, and $-\hat{\alpha}_{\rm M}$.
Since the velocity is of the order of
$u_{\rm ref}\equiv A_{\rm f}/(\nu k_{\rm f}^2)$,
we have defined the fluid and magnetic Reynolds numbers as
$\Rey=u_{\rm ref}/(\nu k_{\rm f})$ and
$R_{\rm m}=u_{\rm ref}/(\eta k_{\rm f})$, respectively.
For all runs we have assumed $\Rey=1$ and $B_0=u_{\rm ref}$.

Again, we see quite clearly the approach of $-\hat{\alpha}_{\rm M}$ toward
$\hat{\alpha}_{\rm F}$, so as to make their sum diminish toward $\alpha$
with increasing values of $R_{\rm m}$.
For $R_{\rm m}<1$ the numerical data agree well with the analytic ones,
whilst for $R_{\rm m}>1$ the numerical values for all alphas lie below the
analytic ones (not shown here).
In particular, for $R_{\rm m}>1$ the value of $\hat{\alpha}_{\rm K}$, based on
the integral in Eq.~(\ref{emfbu1}), begins to exceed the value of $\alpha$.
This apparently signifies the break-down of FOSA.
However, one may expect that the relevant inverse time scales or rates are no
longer governed by just the resistive rate, $\sim \eta k_{\rm f}^2$,
but also by a dynamical rate, $\sim u_{\rm ref} k_{\rm f}$.
This leads to a correction factor, $1/(1+aR_{\rm m})$, where $a\approx1$
is an empirically determined coefficient quantifying the importance of
this correction.
In Fig.~\ref{palptau_vs_Rm_nu2_B2_corr2} we show that both
$\hat{\alpha}_{\rm F}+\hat{\alpha}_{\rm M}$ as well as $\hat{\alpha}_{\rm K}/(1+aR_{\rm m})$
with $a=0.7$ are close to $\alpha$ for $R_{\rm m}\leq30$.
Note that no correction is necessary for $\hat{\alpha}_{\rm F}$ or
$\hat{\alpha}_{\rm M}$, because these quantities are determined by the
momentum equation and hence the viscous time scale.
However, since $\Rey$ is small, no correction is necessary here.
Again, a numerical resolution of $32^3$ meshpoints was used except
for $R_{\rm m}\geq10$, where we used $64^3$ meshpoints.

\section{Time-dependent forcing} 
\label{Section:timedep}

We now consider the case when $\ff$ depends on time, but is ne\-ver\-the\-less
statistically stationary.
In that case, both $\dot{\bb}$ and $\dot{\uu}$ are finite, and hence
also $\overline{\uu\times\dot{\bb}}$ and $\overline{\dot{\uu}\times\bb}$
can in general be finite, even though their sum might vanish in the
statistically steady state.
Later we specialize to one case of particular
interest, when the correlation time of the forcing is small.
This was the case, for example, in the
simulations of \cite{B01} and \cite{BS05b}.
As in Section~\ref{Section:meanemf}, 
we will assume here small magnetic and fluid Reynolds numbers
and neglect the nonlinear terms in the induction and 
momentum equations, but retain the time dependence.
We will also now take a Fourier transform in time and define
\EQ
{\tilde{\uu}}({\bm k},\omega)=
{1\over (2\pi)^4}{\int{\bm u}({\bm x},t)e^{-\ii\kk\cdot\xx + \ii \omega t}\diff{\bm x}}\,\diff t,
\EN
which satisfies the inverse relation 
\EQ
{\uu}({\bm x},t)={\int{{\tilde{\uu}}({\bm k},\omega)
e^{\ii\kk\cdot\xx - \ii \omega t}\diff{\bm k} \,\diff \omega}}.
\EN
In Fourier space, Eqs~(\ref{bintu})  and (\ref{uintbf1}) become
\EQ
\label{bintut}
(-\ii\omega + \eta k^2)\tilbb_{i}({\bm k},\omega)=
(\ii\kk\cdot\BB_0)\,\tiluu_{i}(\kk,\omega),
\EN
\EQ
\label{uintbf1t}
(-\ii\omega + \nu k^2)\tiluu_{i}(\kk,\omega)=
(\ii\kk\cdot\BB_0)\,\tilbb_{i}(\kk,\omega)+\tilff_{i}(\kk,\omega).
\EN
In order to simplify the writing of the equations below, it is convenient
to define complex frequencies
\EQ
\Gamma_\eta = -\ii\omega + \eta k^2,\quad
\Gamma_\nu =-\ii\omega + \nu k^2.
\EN
As before, we can solve equations \eq{bintut} and \eq{uintbf1t} simultaneously
to express $\tilde{\uu}$ and $\tilde{\bb}$ completely in terms of $\tilde{\flucf}$,
\EQ
\label{uintft}
{\tiluu_{i}}({\bm k},\omega)=
\frac{\tilff_i(\kk,\omega)}{\Gamma_\nu + ({\BB_0\cdot\kk})^2/\Gamma_\eta},
\EN
\EQ
\label{bintft}
{\tilbb_i}(\kk,\omega)=
\frac{\tilff_i(\kk,\omega)}{\Gamma_\nu +({\BB_0\cdot\kk})^2/\Gamma_\eta }
\,\frac{\ii\kk\cdot\BB_0}{\Gamma_\eta }.
\EN
We can use these solutions to calculate $\meanEMF$. For getting
an explicit expression, we also need the force correlation function in
Fourier space.
For isotropic, homogeneous and statistically stationary forcing, this is given by
\EQ
\label{ffcort}
\overline{{\tilff_{j}}({\pp},\omega)\;{\tilff_{k}}({\qq},\omega')} =
\delta^3({\pp+\qq})\delta(\omega+\omega')\fjk(\qq,\omega),
\EN
where we can still take $\fjk$ to be of the form given by Eq.~(\ref{fjk}),
with spectral functions now changed to say, a frequency dependent 
$\bar\Phi(k,\omega)$ and $\bar\chi(k,\omega)$.
 
Note that in the limit where the correlation time of the forcing function
is short (delta-correlated in time) the
Fourier space spectral function $\fjk$ is nearly 
independent of the frequency $\omega$.
However if one evaluates the helicity of the forcing, one gets
\EQ
\int \bar \chi(k,\omega)\;\diff{k}\,\diff\omega
=\overline{\ff\cdot(\nab\times\ff)}
\equiv H_{\rm f},
\EN
where the $k$ integral is from $0$ to $\infty$, while the 
$\omega$ integration goes from $-\infty$ to $+\infty$.
For $\bar\chi$ independent of $\omega$ this would be infinite.
So we still keep a spectral dependence and write $\bar\chi(q,\omega) 
=\chi(q) g(\omega)$, where $g$ is an even function of $\omega$, satisfying
$\int g(\omega) \diff\omega = 1$. 
[The property that $g$ is even
is a consequence of the forcing function being real; see
Eq.~(7.44) of \citep{M78}.]
For a forcing with say a correlation time $\tau$,
$g(\omega)$ will be nearly constant for $\omega\tau \sim 1 $ and 
decay at large $\omega$.
In the limit of small $\tau$ the range
for which $g(\omega)$ is nearly constant will be very large. We will need only 
$g(0) \sim \tau$ in most of what follows.
Note that the other extreme limit of steady forcing corresponds to taking
$g(\omega) \to \delta(\omega)$.
The mean emf can be written as
\EQ
\label{meanemft}
\meanemf_{i}({\bm x},t)=
{\epsilon_{ijk}}{\overline {{\uu}_j{\bb}_k}}
= \int {{\tilde{\meanemf_{i}}({\bm k},\Omega)}\;e^{\ii\kk\cdot\xx - \ii\Omega t}
\diff{\bm k}}\,\diff\Omega,
\EN
where the Fourier transform $\tildemeanEMF$ is given by
\EQ
\label{fsemft}
\tilde{\meanemf_{i}}({\bm k},\Omega)= \epsilon_{ijk} {\int{\ubkqt
\;\diff{\bm q}\,\diff\omega}}.
\EN
We now turn to the calculation of the nonlinear mean emf and the resulting
nonlinear $\alpha$-effect. We focus on Method~A, the direct Method~B and also
Method~C to illustrate the similarities and differences from
the case when the time evolution is neglected.
We also discuss in detail the result of applying a 
$\tau$-approximation type method in the
subsequent section. 

\subsection{Computing $\meanEMF$ from the induction equation }
\label{methat}
As before, we first start from the induction equation to solve
for $\bb$ in terms of $\uu$ using Eq.~(\ref{bintut}), so
\EQ
\label{emfbutdep}
{\tilde{\meanemf_{i}}({\bm k})}=
{\ii}{\epsilon_{ijk}}\;\int \BB_0\cdot\qq
\;\frac{\uukqt}{-\ii\omega + \eta q^2}\;\diff{\bm q} \diff\omega.
\EN
We can then express
$\uu$ in terms of $\ff$ using Eq.~(\ref{uintft}). Substituting
from Eq.~(\ref{ffcort}) for the force correlation
in a time dependent flow, the mean emf in coordinate space is then given
by 
\EQ
\label{emfmetbt}
{\meanemf_{i}({\bm x})}=
\ii{\epsilon_{ijk}}\int {{\BB_{0}\cdot{\bm q}}\over{\Gamma_\eta}}
\frac{F_{jk}(\qq, \omega)}
{|\Gamma_\nu +{(\BB_ 0\cdot\qq)^2}/{\Gamma_\eta}|^2}\;
\diff{\qq}\,\diff\omega.
\EN
As usual, we define $\alpha=\meanEMF\cdot\BB_0/B_0^2$.
Since only the
antisymmetric part of $F_{jk}$ contributes in the above integral, 
we have,
\EQ
\label{alpKhat}
\alpha=
-\int\!\frac{(\hat{\BB}_0\cdot\qq)^2\;\chi(q)\;
g(\omega)\;(\ii\omega+\eta q^2)\;\diff\qq\,\diff\omega}
{4\pi q^4\;\vert\;\Gamma_{\nu}\Gamma_{\eta} 
+ (\BB_0\cdot\qq)^2\;\vert^2\;},
\EN
In the following we refer to this expression for $\alpha$ also as $\hat{\alpha}_{\rm K}$,
since it is seen to arise purely from the ($\nabla^{-2}\uu \times \uu$)-type velocity 
correlation, generalized to the time dependent case.
Note that the denominator of Eq.~(\ref{alpKhat}) is even in $\omega$ and
so is the spectral function $g(\omega)$. Therefore
the term in the above integral, which has an $\ii\omega$ in the numerator,
vanishes on integration over $\omega$ (by symmetry). So the mean emf
is a real quantity as it should be.
Before evaluating the above integral
explicitly, let us ask if we get the same expression for Eq.~(\ref{emfmetbt})
using Methods~B and C, even in the time-dependent case.

\subsection{Computing $\meanEMF$ directly }
\label{methbt}
Let us directly compute 
$\meanEMF={\overline {\uu\times\bb}}$ by
substituting $\uu$ and $\bb$ 
in terms of $\flucf$, using Eqs~(\ref{uintft}) and (\ref{bintft}).
We also substitute from Eq.~(\ref{ffcort}) for the force correlation
in a time dependent flow.
The mean emf in coordinate space is then given by Eq.~(\ref{emfmetbt}),
so we do not repeat it here.

\subsection{$\meanEMF$ from the momentum equation}  

As before we start from the momentum equation,
solve for $\uu$ in terms of $\bb$ and the forcing function $\flucf$, 
and then substitute for $\bb$ in terms of $\flucf$ 
using Eq.~(\ref{bintft}). We particularly wish to examine
if the $(\nabla^{-2}\flucf)\times\bb$-like correlation is
essential for calculating the $\meanEMF$ correctly, even for
the time dependent case. 
Using Eq.~(\ref{uintbf1t}) one can write 
\EQ
\label{uintbf2}
{\tiluu_{i}}({\bm k},\omega)=
\frac{1}{\Gamma_\nu }\left[(\ii\kk\cdot\BB_{0})\,{\tilbb_{i}}({\bm k},\omega)
+{\tilff_{i}}({\bm k},\omega)\right].
\EN
From Eq.~(\ref{fsemft}) the mean emf can then be written as  
\EQA
{\tilde{\meanemf_{i}}({\bm k},\Omega)}&\!=\!&
{\epsilon_{ijk}}{\int {\frac{\fjbkt}{-\ii(\Omega -\omega) + \nu(\kk-\qq)^2}\;
\diff{\bm q} \,\diff\omega}} 
\nonumber \\
&&\!\!+{\ii}{\epsilon_{ijk}}
{\int \frac{\BB_0\cdot({\bm{k-q}})}{-\ii (\Omega-\omega) + \nu(\kk-\qq)^2}}
\nonumber \\
&& \qquad \times \; {\bjbkt} \;\diff{\bm q} \,\diff\omega.
\label{metCt}
\ENA
Here the first term involves the $(\nabla^{-2}\flucf)\times\bb$-like
correlation, generalized to the time-dependent case.
Substituting $\tilde{\bb}$ in terms of $\tilde{\flucf}$ from Eq.~(\ref{bintft})
and integrating over the delta functions in wavenumbers and frequencies, 
the mean emf in coordinate space can be expressed as
\EQA
\label{emfmct}
{\meanemf_{i}({\bm x})}&=&
\ii{\epsilon_{ijk}}\int\!\frac{\BB_0\cdot\qq}{\Gamma_\eta }
\frac{F_{jk}(\qq,\omega)}{\vert\Gamma_\nu\vert^2 \;
\left[1+\bar N\right]}\,\diff{\qq}\,\diff\omega \\ \nonumber
&&-\ii{\epsilon_{ijk}}\int\!\frac{\BB_0\cdot\qq}{\Gamma_\eta}
\frac{F_{jk}(\qq,\omega)}{\vert\Gamma_\nu \vert^2\;\left[1+\bar N\right]}\,
\frac{\bar N^*}{1+\bar N^*}\,\diff{\qq}\,\diff\omega.
\ENA
Here we have defined $\bar N = ({\BB_0\cdot\qq})^2/\Gamma_\eta \Gamma_\nu $.
Now the limit of small Lorentz forces corresponds to taking 
$\vert \bar N\vert \ll 1$. Again in this limit the second integral 
vanishes while the first one [i.e.\ the generalized
$(\nabla^{-2}\flucf)\times\bb$-like correlation in Eq.~(\ref{emfmct}),
goes over to a kinematic $\alpha$-effect.
In fact, this part of the $\alpha$-effect can be obtained from
Eqs~(\ref{uintft}) and (\ref{bintft}) by neglecting the
Lorentz force in the expression for ${\tiluu_{i}}({\bm k},\omega)$.

On adding the two terms in Eq.~(\ref{emfmct}),
the mean emf turns out to be 
identical to Eq.~(\ref{emfmetbt}), obtained when starting from
the induction equation or in the direct Method.
Therefore, one could either compute the mean emf starting from the
induction equation, or directly, or from the momentum equation,
by the addition of a generalized $(\nabla^{-2}\flucf)\times\bb$-like 
correlation and a purely magnetic correlation.

We can again define $\hat{\alpha}_{\rm F}$ and $\hat{\alpha}_{\rm M}$
for the time dependent forcing, from the first and the second terms on the
r.h.s.\ of Eq.~(\ref{emfmct}) respectively. We have
\EQ
\label{alpFhat}
\hat{\alpha}_{\rm F} =
-\int\!\frac{(\hat{\BB}_0\cdot\qq)^2\;\chi(q)\;
g(\omega)\left[\Gamma_{\nu}^*\Gamma_{\eta}^* + (\BB_0\cdot\qq)^2\right]
\;\diff\qq\,\diff\omega}
{4\pi q^4\;\Gamma_{\nu}^*\;\vert\;\Gamma_{\nu}\Gamma_{\eta}
+ (\BB_0\cdot\qq)^2\;\vert^2\;},
\EN
\EQ
\label{alpMhat}
\hat{\alpha}_{\rm M} =
\int\!\frac{B_0^2\;(\hat{\BB}_0\cdot\qq)^4\;\chi(q)\;
g(\omega)\;\diff\qq\,\diff\omega}
{4\pi q^4\;\Gamma_{\nu}^*\;\vert\;\Gamma_{\nu}\Gamma_{\eta}
+ (\BB_0\cdot\qq)^2\;\vert^2\;}
\EN
In terms of the total $\alpha$, 
we have $\alpha=\hat{\alpha}_{\rm F} + \hat{\alpha}_{\rm M}$.
It is now explicitly apparent that in the time dependent case,
$\alpha=\hat{\alpha}_{\rm K}=\hat{\alpha}_{\rm F}+\hat{\alpha}_{\rm M}$
in agreement with what is obtained for steady forcing. Further, in the limit
of steady forcing where, $g(\omega) \to \delta(\omega)$, the above
generalized expressions reduce to the $\hat{\alpha}$'s obtained from
Eqs~(\ref{emfmetan}) and (\ref{emfmc}) respectively.

\subsection{The nonlinear $\alpha$-effect}
\label{NonlinAlpTimedep}
Let us return to the explicit computation of the nonlinear $\alpha$-effect 
for the delta-correlated flow.
Solving for
$\alpha=\meanEMF\cdot\BB_0/B_0^2$ 
from Eq.~(\ref{alpKhat}) leads to
\EQ
\label{alpha2t}
\alpha=
-\int \chi(q)\, (\hat{\BB}_0\cdot\hat{\qq})^2 \; (\eta q^2) \; I(\qq) \;
{\diff{\qq}\over4\pi q^2},
\EN
where $I$ is the integral over $\omega$ given by
\EQ
I = \int \frac{g(\omega) \diff\omega}{|\Gamma_\nu \Gamma_\eta +{(\BB_ 0\cdot\qq)^2}|^2}
= \tau \int \frac{\diff\omega}{|\Gamma_\nu \Gamma_\eta +{(\BB_ 0\cdot\qq)^2}|^2}
.
\label{Ieval}
\EN
where the latter expression for $I$ obtains in the limit of a
delta-correlated forcing, where $g(\omega) = \tau$ is almost constant
through the range where the rest of the integrand contributes
significantly.
We now focus on the special case of $\nu=\eta$ for the explicit
evaluation of the above integral. Note that calculating $I$ 
is then straightforward but tedious. We briefly
outline the steps and then quote the result. First the denominator
of Eq.~(\ref{Ieval}) can be expanded and then factorized to give
\EQ
|\Gamma_\nu \Gamma_\eta +{(\BB_ 0\cdot\qq)^2}|^2
=(\omega \!+\! z)(\omega \!+\! z^*)(\omega \!-\! z)(\omega \!-\! z^*),
\label{factorize}
\EN
where $z= (\BB_ 0\cdot\qq) + \ii \nu q^2$.
Then the integral can be rewritten as
\EQ
\intg\! \frac{\tau \diff\omega}{2 (z^{*2} -z^2)\vert z\vert^2}
\left[\!\frac{z}{\omega -z^*}\!-\!\frac{z}{\omega +z^*}\!-\!
\frac{z^*}{\omega -z}\!+\!\frac{z^*}{\omega +z}\!\right]\!.
\EN
Grouping the term having $\omega\!-\!z^*$ with the one having $\omega\!-\!z$ and
likewise the two terms with $\omega\!+\!z^*$ and $\omega\!+\!z$, we get
\EQA
I &=&\intg \frac{\tau \diff\omega}{2 (z^* + z)\vert z\vert^2}\quad\times
\\ \nonumber
&& \left[\frac{-\omega + 2\BB_ 0\cdot\qq}{(\omega -\BB_ 0\cdot\qq)^2 + \nu^2q^4}
+ \frac{\omega + 2\BB_0\cdot\qq}{(\omega +\BB_0\cdot\qq)^2 
+ \nu^2q^4} \; \right].
\ENA 
where the last expression is explicitly real. A change of variables
allows the above integral to be done easily to give%
\footnote{We thank K.-H.\ R\"adler for pointing out to us that this
result can be generalized for $\eta\neq\nu$, to give
$I = \pi\tau/\{(\eta+\nu)q^2[(\BB_0\cdot\qq)^2+\eta\nu q^4]\}$.}
\EQ
I= \frac{\pi}{2}\frac{\tau}{\nu q^2[(\BB_ 0\cdot\qq)^2 +\nu^2q^4]}.
\EN
Substituting $I$ into Eq.~(\ref{alpha2t}), and carrying out
the angular integral over $\mu=\hat{\BB}_0\cdot\hat{\qq}$ then gives
\EQ
\label{alpha3t}
\alpha=-\frac{\pi}{2} 
\tau \int_{0}^{\infty}\frac{\chi(q)}{\nu^2q^4}\,G(\beta)\,\diff{q},
\EN
where $\beta=B_0/\nu q$ for the $\nu=\eta$ case, and
\EQ
G(\beta)= {1\over\beta^2}\left(1- {\tan^{-1}\!\beta\over\beta}\right).
\EN
So for the time dependent delta-correlated flow, 
in the asymptotic limit of large $B_0$, we have
$\alpha \rightarrow B_0^{-2}$.
If we further assume that
the forcing is at a particular wavenumber, $q_0$, and choose 
$\chi(q) = H_{\rm f}\delta(q - q_{0})$, we have 
\EQ
\label{totalphat}
\frac{\alpha}{\alpha_0}=
3\left({B_0\over B_{\rm cr}}\right)^{-2}
\left[1- {\tan^{-1}(B_0/B_{\rm cr})\over B_0/B_{\rm cr}}
\right],
\EN
where we have now defined
\EQ
\alpha_0=-\frac{\pi}{6} \; \frac{\tau H_{\rm f}}{q_0^2B_{\rm cr}^2} ,\quad
B_{\rm cr}=\nu q_{0}.
\label{Defalp0Bcr}
\EN

\subsection{Comparison with $\tau$-approximation}
\label{tauUnsteady}

Note that in the case of time dependent forcing,
considered for example in the
simulations of \cite{B01} and \cite{BS05b}, both $\dot{\bb}$ and
$\dot{\uu}$ are finite, and hence also $\overline{\uu\times\dot{\bb}}$
and $\overline{\dot{\uu}\times\bb}$ can in general be finite, even
though their sum would vanish in the statistically steady state.
So taking a time derivative of $\meanEMF$ and then examining the stationary
situation, could break the degeneracy between $\hat{\alpha}_{\rm K}$ and
$\hat{\alpha}_{\rm F}$ in the kinematic limit, and also lead to novel
insights. Indeed, if one were not able to solve
for $\uu$ and $\bb$ explicitly in terms of the forcing
this would be the practical route to follow.

We now examine the time-dependent case in a manner analogous
to the so-called $\tau$-approximation. As mentioned earlier,
the $\tau$-approximation closures involve invoking a closure whereby triple correlations
which arise during the evaluation of $\partial\meanEMF/\partial t$,
are assumed to provide a damping term proportional to $\meanEMF$ itself.
In the present context there is no need to invoke a closure for the
triple correlations, because
these terms are small for low fluid and magnetic Reynolds numbers.
It turns out that the correct expression for $\meanEMF$ can still be
derived in the same framework, where one evaluates the
$\partial\meanEMF/\partial t$ expression.
Further, it allows us to define a new set of $\alpha$'s
for the time dependent forcing, $\alpha_{\rm K}, \alpha_{\rm M}$ and
$\alpha_{\rm F}$, i.e.\ without hat,
which respectively incorporate the kinetic ($\uu,\uu$), magnetic ($\bb,\bb$)
and the force-field ($\ff,\bb$) correlations (see below). These $\alpha$'s
have properties very similar to those which arise
in the $\tau$-approximation closure for large $\Rm$ systems.

For example, we showed above that for steady forcing, 
the $\nabla^{-2}\ff \times \bb$ correlation is non-zero and essential to
calculate the $\meanEMF$ correctly in Method~C.
We also showed above that such a correlation is important to include
even for a time-dependent forcing, if one starts 
with the explicit solution of the momentum equation for $\tilde{\uu}$.
On the other hand, in a $\tau$-approximation type approach one 
first evaluates the $\partial\meanEMF/\partial t$ expression,
which involves a $\overline{\dot{\uu}\times\bb}$ type correlation,
instead of solving first for $\uu$.
It is then interesting to ask, is the corresponding $(\ff,\bb)$ correlation, 
or the $\alpha_{\rm F}$ term defined below, which arises in the evaluation of
$\overline{\dot{\uu}\times\bb}$,
still non-zero for the time-dependent, delta-correlated forcing? Or does it 
vanish in the $\tau$-approximation type approach, as assumed
in earlier work \citep{BF02,RKR,BS05}?
In particular, does one then recover 
$\alpha=\alpha_{\rm K} + \alpha_{\rm M}$, a 
relation which one obtains in the $\tau$-approximation at large $\Rm$?
We examine these issues in detail below.

We write the time derivative of the emf 
as $\dot{\meanEMF} = \dot{\meanEMF}_{\rm K} + \dot{\meanEMF}_{\rm M}$, where
$\dot{\meanEMF}_{\rm K}= \overline{\uu\times\dot{\bb}}$ and $\dot{\meanEMF}_{\rm M} = 
\overline{\dot{\uu}\times\bb}$. From the induction equation for $\bb$ and the
momentum equation for $\uu$, we now have
\EQ
\dot{\meanEMF}_{\rm K} = \overline{\uu\times\dot{\bb}} = \overline{\uu \times \BB_0\cdot\nab\uu} 
+\overline{\uu \times \eta\nab^2{\bb}},
\label{udotb}
\EN
\EQ
\dot{\meanEMF}_{\rm M} = \overline{\dot{\uu}\times\bb} 
= \overline{\flucf \times \bb} + \overline{
\BB_0\cdot\nab{\bb} \times \bb}
+\overline{\nu\nab^2\uu \times \bb},
\label{dotub}
\EN
where the perturbed pressure term vanishes for divergence free forcing.
We can evaluate each of these terms in Fourier space, 
since we have the Fourier space solutions for both 
$\tilde{\uu}$ and $\tilde{\bb}$ completely in terms of $\tilde{\flucf}$.

In the time dependent case, we define by analogy to Eq.~(\ref{EEdef}),
\EQ
\label{emftdep}
\EE(\kk,\qq, \Omega, \omega) = 
\overline{\tilde{\uu}(\kk-\qq,\Omega-\omega)\times\tilde{\bb}(\qq,\omega)},
\EN
so that the emf in coordinate space is,
\EQ
\meanEMF = \int {{{\EE}(\kk,\qq,\Omega,\omega)}\;
e^{\ii\kk\cdot\xx - \ii\Omega t}\diff{\bm k}}\,\diff{\bm q}\,\diff\Omega\,
\diff\omega . 
\label{emftotal}
\EN
We also define, analogous to Eqs~(\ref{ubdottindep}) and (\ref{udotbtindep}),
\EQ
{\EE}_{{\rm K}}^{(t)}(\kk,\qq,\Omega,\omega)=
\overline{\tilde{\uu}(\kk-\qq, \Omega-\omega)\times[-\ii\omega{\tilde{\bb}}
(\qq,\omega)]} ,
\label{ubdottidep}
\EN
\EQ
{\EE}_{\rm M}^{(t)}(\kk,\qq,\Omega,\omega) =
\overline{[-\ii(\Omega-\omega){\tilde{\uu}}
(\kk-\qq, \Omega-\omega)]\times\tilde{\bb}(\qq,\omega)} .
\label{udotbtdep}
\EN
Note that ${\EE}_{{\rm K}}^{(t)} + {\EE}_{\rm M}^{(t)} = -\ii\Omega \EE$.
Using the induction and momentum equations, we have explicitly,
\EQA
\label{emfKdottdep}
{\EE}_{\rm K}^{(t)}&=&\ii{\qq\cdot\BB_0}\,
[\overline{\tilde{\uu}(\kk-\qq, \Omega-\omega)\times\tilde{\uu}(\qq,\omega)}]
-\eta q^2 \EE
\\
\label{emfMdottdep}
{\EE}_{\rm M}^{(t)} &=&\ii(\kk-\qq)\cdot\BB_0\,
[\overline{\tilde{\bb}(\kk-\qq,\Omega-\omega)\times\tilde{\bb}(\qq,\omega)}]
\nonumber \\
&+&\overline{\tilde{\ff}(\kk-\qq,\Omega-\omega)\times\tilde{\bb}(\qq,\omega)}
-\nu (\kk-\qq)^2 \EE
\ENA
We add Eq.~(\ref{emfKdottdep}) and Eq.~(\ref{emfMdottdep}), to write
\EQA
-\ii\Omega{\EE} + \frac{\EE}{\tau_{\rm eff}}
&=& \ii{\qq\cdot\BB_0}\Phi(\tilde{\uu},\tilde{\uu})
+ \ii(\kk-\qq)\cdot\BB_0\Phi(\tilde{\bb},\tilde{\bb}) \nonumber \\
&& \quad + \Phi(\tilde{\ff},\tilde{\bb}),
\label{newMTA}
\ENA
where we have defined, for any pair of vector fields $\ff_1$ and $\ff_2$,
\EQ
\Phi(\tilde{\ff}_1,\tilde{\ff}_2) =
\overline{\tilde{\ff_1}(\kk-\qq, \Omega-\omega)\times\tilde{\ff_2}(\qq,\omega)},
\label{defPhi}
\EN
and $\tau_{\rm eff}^{-1} = \eta q^2 + \nu (\kk-\qq)^2$.
We note in passing that the above equation is similar to 
the corresponding equation which obtains under $\tau$-approximation in the large
Reynolds number case, except that $\tau_{\rm eff}$ would then correspond
to a relaxation time for triple correlations \citep[cf.][]{BS05}.
So we have,
\EQ
\EE =
{\tau_{\rm eff}^*}
\left[\ii{\qq\cdot\BB_0}\Phi(\tilde{\uu},\tilde{\uu})
+ \ii(\kk-\qq)\cdot\BB_0\Phi(\tilde{\bb},\tilde{\bb})
+ \Phi(\tilde{\ff},\tilde{\bb}) \right].
\label{newMTAstat}
\EN
where we define 
$\tau_{\rm eff}^*=\tau_{\rm eff}/[1-\ii\Omega\;\tau_{\rm eff}]$.\\
Let us define $\alpha = (\meanEMF\cdot\BB_0)/B_0^2$ as before. Then
in coordinate space,
\EQ
\alpha = \alpha_{\rm K} + \alpha_{\rm M} + \alpha_{\rm F},
\EN
where
\EQ
\alpha_{\rm K} = 
\int \ii{\qq\cdot\hat{\BB}_0} \; \Phi(\tilde{\uu},\tilde{\uu})
\cdot\hat{\BB}_0 \; \tau_{\rm eff}^*\;
e^{\ii\kk\cdot\xx - \ii\Omega t}\diff{\bm k}\,\diff{\bm q}\,\diff\Omega\,
\diff\omega,
\label{alK1}
\EN
\EQ
\alpha_{\rm M} = 
\int \ii{\qq\cdot\hat{\BB}_0}\; \Phi(\tilde{\bb},\tilde{\bb})
\cdot\hat{\BB}_0\; \tau_{\rm eff}^*\;
e^{\ii\kk\cdot\xx - \ii\Omega t}\diff{\bm k}\,\diff{\bm q}\,\diff\Omega\,
\diff\omega,
\label{alM1}
\EN
\EQ
\alpha_{\rm F} = 
\int \ii{\qq\cdot\hat{\BB}_0} \; \Phi(\tilde{\ff},\tilde{\bb})
\cdot\hat{\BB}_0\; \tau_{\rm eff}^*\;
e^{\ii\kk\cdot\xx - \ii\Omega t}\diff{\bm k}\,\diff{\bm q}\,\diff\Omega\,
\diff\omega
\label{alF1}
\EN
correspond respectively to the terms containing  the 
$(\uu,\uu)$, $(\bb,\bb)$ and $(\ff,\bb)$ correlations
on the RHS of Eq.~(\ref{newMTAstat}). 

Substituting $\tilde{\uu}$ and $\tilde{\bb}$ in 
terms of $\tilde{\flucf}$ from Eq.~(\ref{uintft}) and (\ref{bintft}), and
integrating over the delta functions in wavenumbers
and frequencies which arises in taking the $(\tilde{\flucf},\tilde{\flucf})$
correlations, we then have in the coordinate space,
\EQ
\label{alK2}
\alpha_{\rm K} =
-\int\!\frac{(\hat{\BB}_0\cdot\qq)^2\;\chi(q)\;
g(\omega)\;\vert\;\Gamma_{\eta}\;\vert^2\;\diff\qq\,\diff\omega}
{4\pi q^4\;\vert\;\Gamma_{\nu}\Gamma_{\eta} 
+ (\BB_0\cdot\qq)^2\;\vert^2\;(\eta+\nu)q^2},
\EN
\EQ
\label{alM2}
\alpha_{\rm M} =
\int\!\frac{B_0^2 (\hat{\BB}_0\cdot\qq)^4\;\chi(q)\;
g(\omega)\;\diff\qq\,\diff\omega}
{4\pi q^4\;\vert\;\Gamma_{\nu}\Gamma_{\eta} 
+ (\BB_0\cdot\qq)^2\;\vert^2\;(\eta+\nu)q^2},
\EN
\EQ
\label{alF2}
\alpha_{\rm F} =
-\!\int\!\frac{(\hat{\BB}_0\cdot\qq)^2\,\chi(q)\,
g(\omega)\left[\Gamma_{\nu}^*\Gamma_{\eta}^* + (\BB_0\cdot\qq)^2\right]
\diff\qq\,\diff\omega}
{4\pi q^4\;\vert\;\Gamma_{\nu}\Gamma_{\eta} 
+ (\BB_0\cdot\qq)^2\;\vert^2\;(\eta+\nu)q^2}.
\label{alf2}
\EN
On adding Eqs~(\ref{alK2}), (\ref{alM2}) \& (\ref{alF2}), the expression
for $\alpha$ is
\EQ
\label{alcomb}
\alpha = 
-\int\!\frac{(\hat{\BB}_0\cdot\qq)^2\;\chi(q)\;
g(\omega)\;I_\alpha}
{4\pi q^4\;(\eta+\nu)q^2}\;\diff\qq\,\diff\omega,
\EN
where 
\EQ
\label{ieq}
I_\alpha = 
\frac{\vert\;\Gamma_{\eta}\;\vert^2 - (\BB_0\cdot\qq)^2 
+ \Gamma_{\nu}^*\Gamma_{\eta}^* + (\BB_0\cdot\qq)^2}
{\vert\;\Gamma_{\nu}\Gamma_{\eta} 
+ (\BB_0\cdot\qq)^2\;\vert^2}.
\EN
The numerator of this integrand can be simplified to give\\
$\Gamma_{\eta}^*[\eta q^2 + \nu q^2]$ such that,
\EQ
\label{altau}
\alpha = 
-\frac{1}{B_0^2}\!\int\!\frac{(\BB_0\cdot\qq)^2\;\chi(q)\;
g(\omega)\;\Gamma_{\eta}^*}{4\pi q^4\;\vert\;\Gamma_{\nu}\Gamma_{\eta} 
+ (\BB_0\cdot\qq)^2\;\vert^2}\;\diff\qq\,\diff\omega.
\EN
It is apparent from Eq.~(\ref{altau}) that the expression
for $\alpha$ turns out to be the same as in Eq.~(\ref{alpKhat}).
Therefore, the $\tau$-approximation type treatment also gives the same $\alpha$
as the other methods.

It is interesting to consider what happens to
the various $\alpha$'s defined in Eqs~(\ref{alK2})--(\ref{alcomb}),
in the limit of steady forcing, where 
$g(\omega) \to \delta(\omega)$.
It is straightforward to check that in this steady forcing limit
$\alpha$ defined in Eq.~(\ref{altau}) goes over exactly to
the total $\alpha = \alpha^{\rm(S)}$ given by the steady state expression in
Eq.~(\ref{alpha2}) of Section 3.5.
Also, in the steady forcing limit, we get
\EQA
\label{alpKrel}
\alpha_{\rm K}&\to&[\eta/(\eta+\nu)] \hat{\alpha}_{\rm K}
= [\eta/(\eta+\nu)] \alpha^{\rm(S)},\\
\alpha_{\rm M}&\to&[\nu/(\eta+\nu)] \hat{\alpha}_{\rm M},\\
\label{alpFrel}
\alpha_{\rm F}&\to&[\nu/(\eta+\nu)] \hat{\alpha}_{\rm F}.
\ENA
In this limit one has therefore
\EQ
\alpha_{\rm M} + \alpha_{\rm F} = [\nu/(\eta+\nu)](\hat{\alpha}_{\rm M}+ 
\hat{\alpha}_{\rm F}) = [\nu/(\eta+\nu)]\alpha^{\rm(S)}
\EN
and so, once again,
\EQ
\alpha_{\rm K} + \alpha_{\rm M} + \alpha_{\rm F} = \alpha^{\rm(S)},
\EN
as expected. 
It should be emphasized, however, that 
for a general time dependent forcing, 
there is no simple relation of the form given by
the expressions (\ref{alpKrel})--(\ref{alpFrel}).

Let us consider the case of delta-correlated forcing now in more detail. 
It is of interest to check if the $\alpha_{\rm F}$ term
contributes in the $\tau$-approximation type closures,
as it does in the time-independent case.
We have from Eq.~(\ref{alf2}), 
\EQ
\alpha_{\rm F} =
-\int \frac{\chi(q)\, (\hat{\BB}_0\cdot\hat{\qq})^2}{(\eta+\nu)q^2} 
\,I_{\rm F}(\qq)\,{\diff{\qq}\over4\pi q^2},
\EN
where $I_{\rm F}$ is the integral over $\omega$ given by
\EQ
I_{\rm F} = \int \frac{\Gamma_\nu^* \Gamma_\eta^* +(\BB_ 0\cdot\qq)^2 }
{|\Gamma_\nu \Gamma_\eta +{(\BB_ 0\cdot\qq)^2}|^2} \; g(\omega)\,\diff\omega.
\label{IevalF}
\EN
Here we can simplify
$\Gamma_\nu^* \Gamma_\eta^* +(\BB_ 0\cdot\qq)^2
= -\omega^2 +\nu\eta q^4 + (\BB_ 0\cdot\qq)^2 +\ii\omega(\eta q^2 + \nu q^2)$.
The integral over the term odd in $\omega$ again vanishes, leaving
again a real $I_{\rm F}$,
\EQ
I_{\rm F} = \int \frac{ -\omega^2 +\nu\eta q^4 +(\BB_ 0\cdot\qq)^2 }
{|\Gamma_\nu \Gamma_\eta +{(\BB_ 0\cdot\qq)^2}|^2} \; g(\omega)\,\diff\omega.
\label{IevalF2}
\EN 
Let us focus on the case $\eta=\nu$ as before.
We rewrite the numerator using the identity
$-\omega^2 +\nu\eta q^4 +(\BB_ 0\cdot\qq)^2
= -(\omega + z)(\omega-z^*) + \omega(z-z^*)$, so we have
\EQ
I_{\rm F} = \int \frac{-(\omega + z)(\omega-z^*)+\omega(z-z^*)}
{|\Gamma_\nu \Gamma_\eta +{(\BB_ 0\cdot\qq)^2}|^2} \; g(\omega)\,\diff\omega.
\label{IevalF3}
\EN
The second term in the numerator of Eq.~(\ref{IevalF3}) does not
contribute to the integral, since it is 
odd in $\omega$, while the denominator is even. To simplify the
integral further we use the identity in Eq.~(\ref{factorize})
for its denominator, giving
\EQA
&I_{\rm F}&\!\! = -\intg \frac{ g(\omega)\,\diff\omega}{(\omega + z^*)(\omega-z) } \\
&=& - \intg \frac{ g(\omega)\,\diff\omega}{z+z^*}
\left[ \frac{1}{\omega-z} - \frac{1}{\omega + z^*}\right]
\nonumber \\
&=& - \intg \frac{ g(\omega)\,\diff\omega}{z+z^*}
\left[ \frac{(\omega -x) +\ii y}{(\omega -x)^2 + y^2}
- \frac{(\omega +x) +\ii y}{(\omega +x)^2 + y^2} \right]\!,
\nonumber
\ENA
where we have defined $x=\BB_0\cdot\qq$ and $y=\nu q^2$, which are the
real and imaginary parts respectively of $z$.
Now changing variables to $u=\omega - x$ in the first term
and $u=\omega + x$ in the second term we see that
\EQ
I_{\rm F} = - \intg \frac{u +\ii y}{z+z^*}\;
\frac{g(u + x) - g(u-x)}{u^2 + y^2} \; \diff u = 0.
\label{if2}
\EN
Note that $I_{\rm F} \to 0$ in the limit of a
delta-correlated forcing, where $g(\omega) = \tau$ is almost constant
through the range where the rest of the integrand contributes
significantly; that is $g(u+x) = g(u-x) = \tau$ where the
integrand contributes significantly, while $g(u+x) \to 0$ and $g(u-x) \to 0$
at large $u$. This can also be checked by doing the integral
for $I_{\rm F}$ numerically.
So, interestingly, $\alpha_{\rm F}=0$. Thus, for a forcing which
is random and delta-correlated in time, there is no contribution
from the $\overline{\ff \times \bb}$ type correlation!
Thus, $\alpha$ is the sum of just two terms, a kinetic and a magnetic
contribution which can be shown explicitly as follows. 

\begin{figure} \begin{center}
\includegraphics[width=\columnwidth]{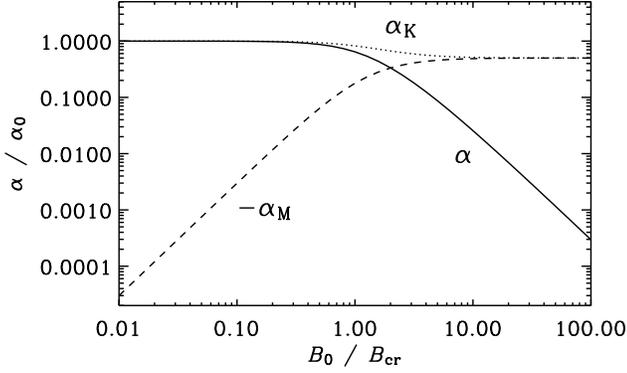}
\end{center}\caption[]{Variation of $\alpha$, $\alpha_{\rm K}$, and
$-\alpha_{\rm M}$ with $B_0$ from the analytical theory using a delta
correlated forcing.
Note that $\alpha=\alpha_{\rm K}+\alpha_{\rm M}$ and $\alpha_{\rm F}=0$.}
\label{analytdc_vs_B0}
\end{figure}

We note that Eq.~(\ref{alK2}) can be expressed as
\EQ
\label{alK3}
\alpha_{\rm K} =
-\frac{1}{4\pi(\eta+\nu)B_0^2}\int
(\BB_0\cdot\qq)^2\; \frac{\chi(q)}{q^6} \;I_{\rm K}(\qq) \;\diff\qq,
\EN
where
\EQA
\label{ik1}
I_{\rm K} &=& \tau\!\int\!\frac{\omega^2 + \eta^2 q^4}
{\vert\;\Gamma_{\nu}\Gamma_{\eta} 
+ (\BB_0\cdot\qq)^2\;\vert^2}\;\diff\omega \nonumber \\
&=& \tau\!\int\!\frac{\omega^2 + \eta^2 q^4 + 
(\BB_0\cdot\qq)^2 - (\BB_0\cdot\qq)^2}
{\vert\;\Gamma_{\nu}\Gamma_{\eta} 
+ (\BB_0\cdot\qq)^2\;\vert^2}\;\diff\omega.
\ENA
As before, we focus on the case $\eta=\nu$ when the numerator
can be simplified as
$\omega^2 + \eta^2 q^4 + 
(\BB_0\cdot\qq)^2 - (\BB_0\cdot\qq)^2 
= (\omega + z)(\omega - z^*)-\omega(z + z^*) - (\BB_0\cdot\qq)^2$.
So we have 
\EQ
\label{ik2}
I_{\rm K} = 
\tau\!\int\!\frac{(\omega + z)(\omega - z^*) - \omega(z + z^*)
- (\BB_0\cdot\qq)^2}
{\vert\;\Gamma_{\nu}\Gamma_{\eta} 
+ (\BB_0\cdot\qq)^2\;\vert^2}\;\diff\omega.
\EN
It is to be noted that the second term in the squared bracket
in Eq.~(\ref{ik2}) does not contribute to the integral. Using the 
identity in Eq.~(\ref{factorize}) for its denominator, we have,
\EQA
\label{ik3}
I_{\rm K} &=&
\tau\!\int\!\frac{\diff\omega}{(\omega - z)(\omega - z^*)} \nonumber \\
&-& \tau\!\int\!\frac{\diff\omega}
{(\omega + z)(\omega + z^*)(\omega - z)(\omega - z^*)} \nonumber \\
&=& \frac{\pi\tau}{\nu q^2} - \frac{\pi\tau}{2\nu q^2}
\;\frac{(\BB_0\cdot\qq)^2}{(\BB_0\cdot\qq)^2 + \nu^2 q^4}.
\ENA
Substituting $I_{\rm K}$ into Eq.~(\ref{alK3}) and carrying out the
angular integral over $\mu=\hat{\BB}_0\cdot\hat{\qq}$ then gives
\EQ
\label{alK_ang}
\alpha_{\rm K} =
-\frac{\pi}{6}\tau\!\int\!\frac{\chi(q)}{\nu^2 q^4}\;\diff\qq
+\frac{\pi}{4}\tau B_0^2\!\int\!\frac{\chi(q)}{\nu^2 q^4}\;H(\beta)\;\diff\qq,
\EN
where $\beta=B_0/\nu q$ for the $\nu=\eta$ case, and
\EQ
H(\beta)= {1\over\beta^2}\left[{1\over3} - {1\over\beta^2}
\left(1- {\tan^{-1}\!\beta\over\beta}\right)\right].
\EN
A similar analysis for Eq.~(\ref{alM2}) yields,
\EQ
\label{alM3}
\alpha_{\rm M} =
\frac{1}{8\pi \nu B_0^2}\int (\BB_0\cdot\qq)^2\;
\frac{\chi(q)}{q^6} \;I_{\rm M}(\qq) \;\diff\qq,
\EN
where 
\EQA
\label{im1}
I_{\rm M} &=& \tau\!\int\!\frac{(\BB_0\cdot\qq)^2\;\diff\omega}
{\vert\;\Gamma_{\nu}\Gamma_{\eta} 
+ (\BB_0\cdot\qq)^2\;\vert^2} \nonumber \\
&=& \frac{\pi\tau}{2\nu q^2}\,
\frac{(\BB_0\cdot\qq)^2}{(\BB_0\cdot\qq)^2 + \nu^2 q^4}.
\ENA
Carrying out the angular integral as earlier gives,
\EQ
\label{alM_ang}
\alpha_{\rm M} =
\frac{\pi}{4}\tau B_0^2\!\int\!\frac{\chi(q)}{\nu^2 q^4}\;H(\beta)\;\diff\qq.
\EN
If we further assume that
the forcing is at a particular wavenumber, $q_0$, and choose 
$\chi(q) = H_{\rm f}\delta(q - q_{0})$, we then have 
\EQ
\label{totalphaKt}
\frac{\alpha_{\rm K}}{\alpha_0}= {1\over2} + 
{3\over2}\left({B_0\over B_{\rm cr}}\right)^{-2}
\left[1- {\tan^{-1}(B_0/B_{\rm cr})\over B_0/B_{\rm cr}}
\right],
\EN
\EQ
\label{totalphaMt}
\frac{\alpha_{\rm M}}{\alpha_0}= -{1\over2} + 
{3\over2}\left({B_0\over B_{\rm cr}}\right)^{-2}
\left[1- {\tan^{-1}(B_0/B_{\rm cr})\over B_0/B_{\rm cr}}
\right],
\EN
where $\alpha_0$ and $B_{\rm cr}$ were defined in Eq.~(\ref{Defalp0Bcr}).
It is explicitly apparent that $\alpha=\alpha_{\rm K}+\alpha_{\rm M}$,
in agreement with Eq.~(\ref{totalphat}).
The result is plotted in Fig.~\ref{analytdc_vs_B0}.
Note also that $\alpha_{\rm F}=0$, as was assumed in
the minimal $\tau$-approximation (MTA) 
type calculations for large fluid and magnetic Reynolds numbers
\citep{BS05}.

It is interesting to note that in the limit $B_0/B_{\rm cr} \gg 1$, 
$\alpha_{\rm K} \to +\alpha_0/2 + O(B_0^{-2})$ and $\alpha_{\rm M} 
\to -\alpha_0/2 + O(B_0^{-2})$ and
so the total $\alpha = \alpha_{\rm K} + \alpha_{\rm M} \to 0$ as
$B_0^{-2}$. This is reminiscent of the kinetic and magnetic $\alpha$'s nearly
cancelling to leave a small residual $\alpha$-effect in EDQNM or MTA type
closures.
It is also interesting to consider the limit when $B_0/B_{\rm cr} \ll 1$.
In this limit $\alpha_{\rm K} \to \alpha_0$ and $\alpha_{\rm M} \to 0$, and
so the net $\alpha$-effect is just the kinetic contribution.
Finally, for any $B_0$ we note that $(\alpha_{\rm K} - \alpha_{\rm M})/\alpha_0 = 1$.

\subsection{Comparison with simulations}

It is appropriate to compare with the simulations of \cite{BS05b}.
We have produced additional results for low fluid and magnetic Reynolds
numbers ($\Rey=R_{\rm m}\approx2\times10^{-2}$).
The forcing consists of helical waves with average wavenumber
$k_{\rm f}/k_1=1.5$.
The resulting values of $\alpha$ are fluctuating strongly, so it is important
to average them in time.
Instead of calculating the full integral expressions, we estimate the
contributions to $\alpha$ from the formulae
$\alpha_{\rm K}=-2\tau\langle u_x u_{z,y}\rangle$,
$\alpha_{\rm M}=2\tau\langle b_x b_{z,y}\rangle$,
$\alpha_{\rm F}=\tau\langle\ff\times\bb\rangle\cdot\BB_0/B_0^2$,
where $\BB_0=(0,B_0,0)$ is the imposed field,
and $\tau^{-1}=(\nu+\eta)k_{\rm f}^2$.

\begin{figure}\begin{center}
\includegraphics[width=\columnwidth]{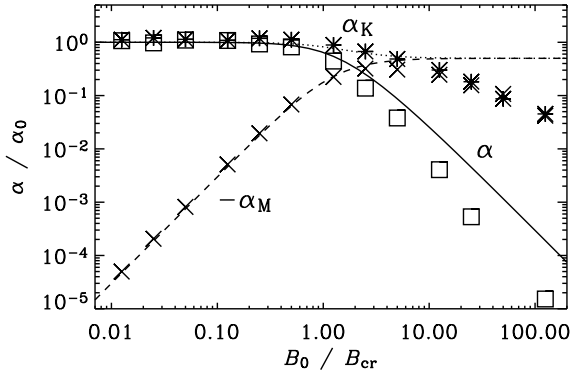}
\end{center}\caption[]{
Variation of $\alpha$, $\alpha_{\rm K}$, and
$-\alpha_{\rm M}$ with $B_0$ for a random flow at low fluid and
magnetic Reynolds numbers ($\Rey=R_{\rm m}\approx2\times10^{-2}$),
compared with the analytic theory predicted for a fully isotropic flow.
Note that the numerically determined values of $\alpha_{\rm K}$ are
smaller and those of $-\alpha_{\rm M}$ larger than the corresponding
analytic values, which is similar to the results for the ABC-flow forcing.
}\label{palpKM_vs_B0}\end{figure}

The result is shown in Fig.~\ref{palpKM_vs_B0} and compared with the
results of the previous section.
In all cases the resulting values of $\alpha_{\rm F}$ are negligibly
small and will not be considered further.
Like in the case of the ABC flow, the numerically estimated values of
$\alpha$ are smaller than the analytic ones.
This might be explicable if for some reason the relevant normalization
in terms of $\alpha_0$ were to depend on $B_0$.
Alternatively, the discrepancy might be due to us using only simplified
expressions instead of the full integral expressions.
However, the important point is that the main contribution to the quenching
comes from the growing contribution of $-\alpha_{\rm M}$ such that
$\alpha_{\rm K}+\alpha_{\rm M}$ is quenched to values much smaller than
$\alpha_{\rm K}$.
The corresponding results in the case of larger $R_{\rm m}$ are given
by \cite{BS05b}.

\section{Discussion} 
\label{discimp}
We have considered here the nonlinear $\alpha$-effect in the
limit of small magnetic and fluid Reynolds numbers, 
for both steady and time-dependent (both for general and delta-correlated)
forcings.
In the limit of low $\Rm$ and $\Rey$, 
one can neglect terms nonlinear in the fluctuating fields and hence 
explicitly solve for the 
small scale magnetic and velocity fields using double FOSA.
We can then calculate the $\alpha$-effect in several different ways.

For both steady and time dependent forcings, one gets similar
results, provided one starts from the explicit solutions
to the induction and momentum equations. Lets us begin
with a summary of the results for the steady forcing case:
To begin with we follow in Method~A, the traditional route of
solving the induction equation for $\bb$ in terms of $\uu$, 
and then calculating the $\alpha$-effect.
For statistically isotropic velocity fields,
this gives $\alpha$ dependent on the helicity of the velocity potential,
as already known from previous work. In addition since we have
an explicit solution for $\uu$ in terms of $\flucf$, one
can relate $\alpha$ directly to the helical part of the force correlation.
In Method~B we solved for $\uu$ and $\bb$ in terms 
of the forcing function $\flucf$ and computed $\meanEMF$ directly. 
This would correspond to what is done when the $\alpha$-effect is
determined from simulations.
However, in general this cannot be done analytically unless one can solve
for the small scale velocity and magnetic field explicitly. 
We get $\alpha$ identical to that obtained in Method~A.

More interesting is Method~C, where one takes the momentum equation as
the starting point, instead of the induction equation.
In the limit of small fluid Reynolds numbers one can solve for
$\uu$ in terms of $\bb$ and hence compute $\meanEMF$. 
This necessarily involves also the $(\nabla^{-2}\flucf)\times\bb$ correlation,
between the forcing and the small-scale magnetic field,
in addition to the $(\nabla^{-2}\jj)\cdot\bb$ (or $\vpa\cdot\bb$)
correlation arising from the
Lorentz force. This second term depends on the helicity
of the small scale magnetic fields.
When the Lorentz force is small, the first term contributes 
to $\alpha$ in a manner closely
related to the usual kinematic alpha-effect, while the
second term contributes negligibly. 
Interestingly, as the Lorentz force gains in importance the first term
is suppressed, while the second term (which has an opposite sign)
gains in importance and cancels the first term, 
to further suppress the total $\alpha$-effect (Fig.~\ref{panalyt_vs_B0}).
This is similar to the suppression of
the kinetic alpha due to the addition of a magnetic alpha
(proportional to the helical part of $\bb$) found in several
closure models \citep{pouq,KR82,GD,BF02,BS05b}.
When one combines the two terms,
the resulting 
$\alpha$-effect is identical to that obtained
from the induction equation (Method~A) using FOSA
or the direct computation of Method~C.
However, it also highlights the fact that 
in this steady case the $(\nabla^{-2}\flucf)\times\bb$-type
correlation does not vanish, and that there is no tendency
for this term to balance the viscous term, as one might have expected.

Finally, the results of Method~D show that the formalism used in the
$\tau$-approximation lead to results that are equivalent to the usual
approach taken in the first order smoothing approximation.
However, this requires that the detailed spectral dependence of the
diffusion operator be retained until the point where the steady
state assumption is made.
The resulting equations are solved for the spectral electromotive force
of the form $\EE(\kk,\qq)$ in \Eq{tau1}.
Other\-wise, one would not recover the correct low conductivity limit,
as shown by \cite{RR07}.
We emphasize that throughout this paper we have understood the
term $\tau$-approximation only in this more generalized sense.

The above results are also obtained for statistically stationary 
but time-dependent forcing. Specifically, we showed that even in the
time dependent case, one gets identical results
for the mean emf if $\alpha$ is computed directly (as in Method~B),
or from the momentum equation, by the addition of a generalized
$(\nabla^{-2}\flucf)\times\bb$-like 
correlation and a purely magnetic correlation. The explicit form of the 
$\alpha$-effect differs between delta-correlated and steady forcing cases. 
In particular, in the limit of large $B_0$, and when $\nu=\eta$, 
we have $\alpha \propto B_0^{-2}$ for delta-correlated
forcing, in contrast to $\alpha \propto B_0^{-3}$, for the case of 
a steady forcing.  
The former result has already been obtained by \cite{FBC99} and \cite{RK00},
both of whom assumed the force--field correlation to vanish.
However, their result was derived under the assumption of large fluid
and magnetic Reynolds numbers.

The major difference between the time dependent and steady forcing cases
arise when one follows Method~D, the formalism used in the
$\tau$-approximation type approaches to computing $\alpha$
in large $\Rm$ systems. 
We recall that in this approach one starts by evaluating 
the time derivative of the emf, and then look at the stationary limit.
We showed that in this limit the $\alpha$-effect can be naturally
written as the sum of 3 terms, $\alpha = \alpha_{\rm K} + \alpha_{\rm M} 
+ \alpha_{\rm F}$, for a general time-dependent forcing. 
Here $\alpha_{\rm K}$ and  $\alpha_{\rm M}$ are the kinetic and
magnetic contributions to $\alpha$ corresponding respectively to the 
terms containing  the
$(\uu,\uu)$ and $(\bb,\bb)$ [see Eqs~(\ref{alK1}) and (\ref{alM1})],
while $\alpha_{\rm F}$ incorporates the $(\ff,\bb)$ correlation;
see Eq.~(\ref{alF1}). Interestingly, we showed that $\alpha_{\rm F}=0$ 
in the approach of Method~D, for delta-correlated forcing, and therefore 
$\alpha=\alpha_{\rm K} + \alpha_{\rm M}$, just the sum of a kinetic and
magnetic terms. We also computed $\alpha_{\rm K}$ and $\alpha_{\rm M}$ 
explicitly for the case $\eta=\nu$. 
In the kinematic limit, $\alpha_{\rm M} \to 0$,
while $\alpha \to \alpha_{\rm K}$.
While in the limit $B_0/B_{\rm cr} \gg 1$,
$\alpha_{\rm K} \to +\alpha_0/2 + O(B_0^{-2})$ and $\alpha_{\rm M}
\to -\alpha_0/2 + O(B_0^{-2})$, 
so that the total $\alpha = \alpha_{\rm K} + \alpha_{\rm M} \to 0$ as
$B_0^{-2}$. This is reminiscent of the kinetic and magnetic $\alpha$'s nearly
cancelling to leave a small residual $\alpha$-effect in EDQNM or
$\tau$-approximation type closures.
The results from employing FOSA and $\tau$-approximation type analysis is
summarized in Table~\ref{results} both for steady and random forcings.

\begin{table} \begin{center}
\caption[]{\label{results}
Summary of the results obtained from FOSA and $\tau$-approximation type analysis for steady
and random forcings.}
\begin{tabular}{lll}
\hline
& steady forcing & delta-correlated forcing \\
\hline
FOSA &$\alpha=\hat{\alpha}_{\rm K}$
     &$\alpha=\hat{\alpha}_{\rm K}$\\
     &$\quad=\hat{\alpha}_{\rm F}+\hat{\alpha}_{\rm M}$
     &$\quad=\hat{\alpha}_{\rm F}+\hat{\alpha}_{\rm M}$ \\
$\tau$-approximation  &$\alpha=\alpha_{\rm K}+\alpha_{\rm F}+\alpha_{\rm M}$  
	&$\alpha=\alpha_{\rm K}+\alpha_{\rm F}+\alpha_{\rm M}$\\
&       &$\quad=\alpha_{\rm K}+\alpha_{\rm M}$ \\
\hline
\end{tabular}
\end{center}
\end{table}

As far as the low Reynolds number case is concerned, our analytic
solutions demonstrate quite clearly that one can look at the nonlinear
$\alpha$-effect in several equivalent ways.
On the one hand, one can express $\alpha$ completely
in terms of the helical properties of the velocity field
(Method~A) as advocated by \citet{proctor} and \citet{RR07}.
At the same time $\alpha$ can be naturally expressed as
a sum of a suppressed kinetic part (first term in Method~C)
and an oppositely signed magnetic part proportional
to the helical part of $\bb$ (second term in Method~C).

Method~D applied to the delta-correlated forcing is particularly revealing.
As here one can explicitly write 
$\alpha = \alpha_{\rm K} + \alpha_{\rm M}$, or as the sum of a kinetic
$\alpha_{\rm K}$, which dominates in the linear regime,
and a magnetic $\alpha_{\rm M}$, which gains in importance
as the field becomes stronger, and cancels $\alpha_{\rm K}$ 
to suppress the net $\alpha$-effect. 
This is similar to the approach that arises from closure models
like EDQNM \citep{pouq} and the $\tau$-approximation
\citep{KMR96,BF,BS05} or the quasilinear models \citep{GD}.
In all these cases the nonlinear $\alpha$-effect, for large $R_{\rm m}$,
is the sum of a kinetic part and an oppositely signed magnetic part.
As noted above, the kinetic $\alpha$-effect is itself suppressed, as
seen in Fig.~\ref{analytdc_vs_B0}, but this happens only for
$B_0>B_{\rm cr}$, and the suppression is milder than the
the strong suppression of the total $\alpha$-effect. 
This is also borne out in the simulations of \cite{BS05b},
where the kinetic part of the $\alpha$-effect is suppressed in a manner
that is independent of the magnetic Reynolds number,
even though the total $\alpha$ is catastrophically suppressed.
Finally, we also have shown that $\alpha_{\rm F}$ defined naturally 
in the approach of Method~D, vanishes for delta-correlated forcing,
as was assumed in derivations of the $\alpha$-effect
in $\tau$-approximation type closures.

In the special case of periodic domains it is now clear that for forced
turbulence at large $R_{\rm m}$ the steady state $\alpha$-effect is
catastrophically quenched \citep{CH96,B01}.
However, the physical cause of this phenomenon was long controversial.
Is it because Lorentz forces cause a suppression of Lagrangian chaos
\citep{CHK96} or is it due to a nonlinear addition to $\alpha$ due to helical
parts of the small scale magnetic field, as is argued here?
The latter alternative is also supported by the excellent
agreement between model calculations and simulations \citep{FB02,BB02}.
Furthermore, the simulations of \cite{BS05b} demonstrate that this
quenching is accompanied by an increase of $-\alpha_{\rm M}$ toward
$\alpha_{\rm K}$, and that this $\alpha_{\rm K}$ itself is unquenched.
Subsequent analysis of their data shows that $\alpha_{\rm K}$ remains
unquenched regardless of whether or not one uses the proper anisotropic
expression (Brandenburg \& Subramanian, unpublished).

Our present results are of course restricted to the case
of small magnetic and fluid Reynolds numbers.
This means that we have not tested any of the actual closure assumptions,
like the $\tau$-approximation.
Such tests have so far only been done numerically \citep{BKM04,BS05}.
Clearly, at large magnetic and fluid Reynolds numbers the
$\tau$-approximation can no longer yield exact results.
Nevertheless, it provides a very practical tool to estimate the
mean field transport coefficients in a way that captures
correctly some of the effects that enter in the case of
large magnetic and fluid Reynolds numbers.
In that respect, it has been quite successful in reproducing the
catastrophic quenching result for periodic domains, as well as 
suggesting ways to alleviate such quenching.

In Section~\ref{MFED} we outlined the conditions under which double FOSA
is valid as being basically the requirement that
$R_{\rm m} \ll 1$ and $\Rey \ll 1$. A subtle point concerns the 
validity of retaining the linear Lorentz force term, $\BB_0\cdot\nabla\bb$,
while neglecting the nonlinear advection, $\uu\cdot\nabla\uu$, even though
nonlinear advection is small compared to the viscous dissipation for $\Rey\ll1$.
This assumption is valid provided $B_0 b/l > u^2/l$,
or, using $b \sim R_{\rm m} B_0$, $B_0^2 R_{\rm m} > u^2$.
(We thank Eric Blackman for pointing this out to us.)
Note that in terms of the critical field $B_{\rm cr} = \sqrt{\nu\eta} q_0$,
which divides the regimes where the Lorentz force is important ($B_0 > B_{\rm cr})$ 
and where it is not ($B_0 < B_{\rm cr}$), this requirement becomes 
$B_0/B_{\rm cr}> \Rey^{1/2}$.
Therefore, for small fluid Reynolds number our assumption of
retaining the linear Lorentz force term
while dropping nonlinear advection is indeed valid for most
regimes of interest for $\alpha$ suppression.
For smaller mean fields, where $B_0/B_{\rm cr} < \Rey^{1/2}$,
in any case the Lorentz force has no impact. The interesting point seems
to be that for low Reynolds number systems, the typical reference mean field
is not the equipartition field $B_0 = u$, but rather
$B_0 = B_{\rm cr} \sim u/(R_{\rm m}\Rey)^{1/2} > u$.

Throughout this work we have adopted an externally imposed body force
to drive the flow.
This is commonly done in many simulations in order to achieve homogeneous
isotropic conditions that are amenable to analytic treatment.
Clearly, this is not the case for many astrophysical flows that are driven
by convection (e.g.\ in stars) or the magneto-rotational instability
(e.g.\ in accretion discs).
Such flows tend to show long-range spatial correlations, which means
that the alpha tensor should really be treated as a integral kernel
\citep[see, e.g.,][]{BS02}.
It is at present unclear whether such more natural forcings are closer
to steady or to random forcing, and how big is the resulting
$(\ff,\bb)$ correlation.
Given that this correlation represents already a linear effect,
it is likely that the $\alpha_{\rm F}$ term can simply be subsumed
into an expression for a modified kinetic $\alpha_{\rm K}$.
If this is the case, we can continue to write
$\alpha \approx \alpha_{\rm K}+\alpha_{\rm M}$ as the sum of a 
mildly suppressed kinetic part depending on the velocity field and
a magnetic part, so that their sum accounts for the tendency toward
catastrophic $\alpha$-quenching in the absence of helicity fluxes.
We recall that such a split is exact in the limit of delta-correlated forcing.

\section{Conclusions}

Our work was motivated in part by the detailed
criticism expressed by \cite{RR07}.
In view of our new results we can now make the following statements
for small magnetic and fluid Reynolds numbers.
Firstly, it is true that the $\hat{\alpha}_{\rm K}$ that is calculated
under FOSA does indeed capture the full nonlinear $\alpha$-effect ---
provided it is based on the actual velocity field.
Secondly, the $\alpha_{\rm K}$ that is calculated in the $\tau$
approximation, is not simply related to $\hat{\alpha}_{\rm K}$,
except in the limit of steady forcing.
Thirdly, in the limit of small magnetic and fluid Reynolds numbers,
both FOSA and the $\tau$-approximation give identical results.
Indeed, all methods of calculating the $\alpha$-effect agree,
as they should, given that the starting equations were the same.
However, the force--field correlation cannot be ignored in general.
The exception is when one analyzes the case of delta-correlated forcing
in a manner akin to the $\tau$-approximation, where
the force--field correlation does vanish and hence
$\alpha_{\rm F} =0$ explicitly.
In this case one can indeed write $\alpha=\alpha_{\rm K} + \alpha_{\rm M}$,
or the sum of a kinetic and magnetic alpha effects.
Furthermore, due to the spatial non-locality of the Greens function
for small magnetic and fluid Reynolds numbers, the $\tau$-approximation
should be carried out at the level of spectral correlation tensors,
as is done here.
Somewhat surprisingly, the delta-correlated forcing case yields an
asymptotic $\alpha \propto B_0^{-2}$ scaling as opposed to the well-known
$\alpha \propto B_0^{-3}$ behaviour for steady forcing.

Although our work is limited to small magnetic and fluid Reynolds numbers,
the calculations of Method~C and Method~D, and its agreement with the results of
Method~A, (for both steady and time-dependent forcings),
do suggest one way of thinking about the effect of
Lorentz forces: they lead to a decrease of $\alpha$ predominantly by
addition of terms proportional to the helical parts of the small
scale magnetic field. Hence getting rid of such small scale magnetic helicity
by corresponding helicity fluxes, may indeed be the way astrophysical
dynamos avoid catastrophic quenching of $\alpha$ to make their dynamos
work efficiently.

\section*{Acknowledgments}
We thank Eric Blackman, Nathan Kleeorin, Karl-Heinz R\"adler,
Igor Rogachevskii, and Anvar Shukurov for valuable comments on the manuscript.
SS would like to thank the Council of Scientific and Industrial Research, 
India for providing financial support.
We acknowledge the Danish Center for Scientific Computing
for granting time on the AMD Opteron Linux cluster in Copenhagen.

\newcommand{\ybook}[3]{ #1, {#2} (#3)}
\newcommand{\yjfm}[3]{ #1, {J.\ Fluid Mech.,} {#2}, #3}
\newcommand{\yprl}[3]{ #1, {Phys.\ Rev.\ Lett.,} {#2}, #3}
\newcommand{\ypre}[3]{ #1, {Phys.\ Rev.\ E,} {#2}, #3}
\newcommand{\yapj}[3]{ #1, {ApJ,} {#2}, #3}
\newcommand{\yan}[3]{ #1, {AN,} {#2}, #3}
\newcommand{\yana}[3]{ #1, {A\&A,} {#2}, #3}
\newcommand{\ygafd}[3]{ #1, {Geophys.\ Astrophys.\ Fluid Dyn.,} {#2}, #3}
\newcommand{\ypf}[3]{ #1, {Phys.\ Fluids,} {#2}, #3}
\newcommand{\yproc}[5]{ #1, in {#3}, ed.\ #4 (#5), #2}
\newcommand{\yjour}[6]{, #6, {#2} {#3} (#1) #4--#5.}

\label{lastpage}
\end{document}